\documentclass[a4paper,showpacs,preprintnumbers,superscriptaddress,aps,prd,nofootinbib,floatfix,10pt]{revtex4-2}
\usepackage{graphicx}
\usepackage{dcolumn}
\usepackage{bm}
\usepackage{amsmath,amssymb,amsfonts}
\usepackage{latexsym}
\usepackage{color}
\usepackage{subcaption}
\usepackage{ulem}
\usepackage{mathrsfs}
\usepackage{braket}
\usepackage{yhmath}
\usepackage{cancel}
\usepackage{tensor}

\usepackage[usestackEOL]{stackengine}
 
\usepackage{multirow}

\usepackage{physics}
\usepackage{titlesec}

\renewcommand{\thesection}{\arabic{section}}
\renewcommand{\thesubsection}{\arabic{subsection}}

\usepackage{hyperref}
\hypersetup{%
setpagesize=false,
 bookmarksnumbered=true,%
 bookmarksopen=true,%
 colorlinks=true,%
 linkcolor=blue,
 citecolor=magenta,
}

\usepackage{graphicx}
\usepackage{dcolumn}
\usepackage{bm}
\usepackage{physics,amsmath,amssymb,amsfonts}
\usepackage{latexsym}
\usepackage{color}
\usepackage{slashed}

\def\nn    {\nonumber}
\DeclareUnicodeCharacter{039B}{\Lambda}

\usepackage{titlesec}

\titleformat{\section}
  {\normalfont\large\bfseries}{\thesection}{1em}{}
\titlespacing*{\section}
  {0pt}{1.ex plus .1ex minus .1ex}{0.ex plus .1ex minus .1ex}

\makeatletter
\renewcommand{\p@subsection}{\thesection.}
\renewcommand{\p@subsubsection}{\thesection.\thesubsection.}
\makeatother

\titleformat{\subsubsection}
  {\normalfont\itshape}{\thesection.\thesubsection.\thesubsubsection}{1em}{}
\titlespacing*{\subsubsection}
  {0pt}{1.ex plus .1ex minus .1ex}{0.ex plus .1ex minus .1ex}

\renewcommand{\thesubsection}{.\arabic{subsection}}

\begin{document}
\allowdisplaybreaks
\flushbottom
\title{\boldmath 
The $R^2$-Higgs inflation: $R^3$ contribution and preheating after ACT and SPT data
}

\author{Tanmoy Modak}
\affiliation{Department of Physical Sciences, Indian Institute of Science Education and Research Berhampur, Berhampur 760003, Odisha, India\\[0.1cm]}

\begin{abstract}
The $R^2$-Higgs inflation is one of the simplest yet best-fit models consistent with Planck data. The higher spectral index $n_s$ recently reported by the combined cosmic microwave background (CMB) data from the Atacama Cosmology Telescope (ACT), South Pole Telescope (SPT), and Planck, along with baryonic acoustic oscillation data from the Dark Energy Spectroscopic Instrument (DESI), disfavors the single-field-like regime of $R^2$-Higgs inflation at approximately $2\sigma$. Following a doubly covariant formalism, we show that the $R^2$-Higgs inflation, when modified by a dimension-six $R^3$ term, can account for the high $n_s$ reported by CMB+BAO. In this regard, we find that preheating may play a pivotal role. We also show that if the nonminimal coupling between the Ricci scalar $R$ and the Higgs field is $\mathcal{O}(10)$, then preheating via the production of Higgs quanta may help explain the reported observations.
\end{abstract}

\maketitle
\vspace{2ex}

\hrule
\vspace{2ex}
 \tableofcontents
\vspace{2ex}
\numberwithin{equation}{section}
\hrule
\setlength{\parskip}{1\baselineskip}
\setlength{\parindent}{0pt}
\vspace{1ex}
\section{Introduction}\label{sec:intro}
The cosmic inflation~\cite{Starobinsky:1980te,Sato:1980yn,Guth:1980zm} offers an elegant solution 
to the so-called flatness, horizon, and exotic relic problems. The quantum fluctuations generated during
inflation later transformed into density perturbations, which subsequently formed the large-scale structures
such as the CMB anisotropies. The features predicted by the inflationary paradigm, such as the acoustic peaks 
in the CMB and the nearly scale-invariant adiabatic primordial power spectrum, are in excellent agreement with 
the WMAP~\cite{WMAP:2012nax} and Planck~\cite{Planck:2018jri} data. While several predictions of the inflationary 
paradigm are in good agreement, the “last missing piece” of cosmic inflation, i.e., the primordial gravitational 
waves generated during inflation, is yet to be discovered. Ongoing experiments~\cite{BICEP:2021xfz,SimonsObservatory:2018koc} 
are actively searching for this last missing piece which are believed to be engraved in the CMB $B$-mode polarization, with future 
missions proposed~\cite{LiteBIRD:2022cnt}.

The spectral index $n_s$ also provides a powerful test for inflationary dynamics. Indeed, several inflationary models have been ruled out based on their predicted values of $n_s$ when compared to the observed $1\sigma$ and $2\sigma$ contours in the $n_s$ vs. $r$ plane ($r$ being the tensor-to-scalar ratio) of the Planck 2018 data~\cite{Planck:2018jri}. Among other inflationary models, the Starobinsky or $R^2$ inflation~\cite{Starobinsky:1980te,Starobinsky:1983zz,Vilenkin:1985md,Mijic:1986iv,Maeda:1987xf} was one of the best-fit models and lay within the $\sim1\sigma$ range in the $n_s$ vs. $r$ plane of the Planck data~\cite{Planck:2018jri}. The marginalized $n_s$ from the ACT~\cite{ACT:2025fju}, SPT~\cite{SPT-3G:2025bzu}, and the combined Planck+ACT+SPT (CMB-SPA) data~\cite{SPT-3G:2025bzu} are still consistent with each other and found to be still on the lower side in $\Lambda \rm{CDM}$ model~\footnote{The marginalised $n_s$ values within $\Lambda \rm{CDM}$ model from different CMB experiments and their combinations are as follows-- Planck: $0.9657 \pm 0.0040$, SPT-3G D1: $0.951 \pm 0.011$, 
ACT DR6: $0.9682 \pm 0.0069$, SPT+ACT: $0.9671 \pm 0.0058$, SPT+Planck: $0.9636 \pm 0.0035$ and combination of Planck+SPT+ACT i.e. CMB-SPA:
$0.9684 \pm 0.0030$ respectively~\cite{SPT-3G:2025bzu}. Note that, all these experiments considered both $T$, $E$ polarization data and lensing reconstruction and,
the prior of $\tau_{\rm reio}$ is taken same as in Planck PR4. The $n_s$ from the joint SPT-3G D1 + DESI is $0.949 \pm 0.012$, which is actually much lower than
both P-ACT-LB2 $n_s = 0.9752 \pm 0.0030$~\cite{ACT:2025fju} and  from CMB-SPA+DESI $n_s=0.9728 \pm 0.0027$~\cite{SPT-3G:2025bzu}.}. However, the $n_s$ is shifted towards a higher value in the $\Lambda \rm{CDM}$ model if one combines CMB with the BAO data from the DESI-DR2 collaboration~\cite{DESI:2025zgx}. This higher $n_s$ is separately reported by both the ACT~\cite{ACT:2025fju} and SPT~\cite{SPT-3G:2025bzu} collaborations; the ACT collaboration found $n_s = 0.9752 \pm 0.0030$ by combining the CMB data from ACT, Planck (with lensing), and BAO data from DESI-DR2 (denoted as P-ACT-LB2)~\cite{ACT:2025fju}, whereas the SPT collaboration reported $n_s=0.9728 \pm 0.0027$ by combining data from SPT, ACT, Planck, and DESI-DR2 (denoted as CMB-SPA+DESI)~\cite{SPT-3G:2025bzu}. This high $n_s$ from CMB+BAO excludes pure $R^2$ inflation at $>2\sigma$~\cite{ACT:2025fju,SPT-3G:2025bzu}.

The $R^2$ inflation model, while successful in many aspects, is not fully realistic. Following the discovery of the Higgs boson at the Large Hadron Collider, incorporating the Higgs field into inflationary dynamics has become essential. This model, i.e., the baseline Starobinsky model extended with the Higgs field and a coupling between $R$ and the Higgs field, dubbed the $R^2$-Higgs inflation~\cite{Salvio:2015kka,Ema:2017rqn,Pi:2017gih,Gorbunov:2018llf,Gundhi:2018wyz,He:2018mgb,He:2018gyf,Cheong:2019vzl,Canko:2019mud,He:2020ivk,He:2020qcb,Pineda:2024prs} (see also Refs.\cite{Copeland:2013vva,Saltas:2015vsc,delaCruz-Dombriz:2016bjj,Odintsov:2020ilr}), has similar inflationary predictions as $R^2$ inflation and is also a best-fit model to Planck data~\cite{Planck:2018jri}. Like $R^2$ inflation, the $R^2$-Higgs inflation is also a best-fit model to Planck data~\cite{Planck:2018jri}, but is now in tension with the measured high $n_s$ from CMB+BAO~\cite{ACT:2025fju,SPT-3G:2025bzu}, at least for the majority of the single-field-like regime. Several attempts have already been made to reconcile both $R^2$ inflation, Higgs inflation and $R^2$-Higgs inflation with the observed high $n_s$~\cite{Gialamas:2025kef,Drees:2025ngb,Zharov:2025evb,Liu:2025qca,Haque:2025uis,Yogesh:2025wak,Addazi:2025qra,Kallosh:2025ijd,Gialamas:2025ofz,Saini:2025jlc,Wolf:2025ecy,Wang:2025dbj,Piva:2025cqi,SidikRisdianto:2025qvk,Zharov:2025zjg,Ferreira:2025lrd,Ellis:2025ieh,Oikonomou:2025htz}.

The inflationary observables, such as $n_s$ and $r$, can receive significant modifications if one adds a dimension-six $R^3$ term to the action of $R^2$-Higgs inflation~\cite{Modak:2022gol,Lee:2023wdm}. Such curvature modifications are motivated not only by $f(R)$ theories of gravity but also from a purely phenomenological point of view~\cite{Saidov:2010wx,Huang:2013hsb,Huang:2013hsb,Motohashi:2014tra,Asaka:2015vza,Bamba:2015uma,Miranda:2017juz,Cheong:2020rao,Rodrigues-da-Silva:2021jab,Ivanov:2021chn,Koshelev:2022olc,Shtanov:2022pdx,Wang:2023hsb,Kim:2025dyi}(see also Ref.~\cite{Sebastiani:2013eqa, Odintsov:2017hbk, Nojiri:2017ncd}). In this paper, we study the implications of the $R^3$ term in $R^2$-Higgs inflation to account for the observed high $n_s$ (see Refs.~\cite{Gialamas:2025ofz,Addazi:2025qra} for a similar discussion). We show that the $R^3$ term can indeed account for the observed $n_s$, as well as the amplitude of the scalar power spectrum, as reported by the ACT~\cite{ACT:2025fju} and SPT~\cite{SPT-3G:2025bzu} collaborations.
We adopt a doubly covariant formalism for our analysis, which is well-suited for the two-field model under consideration. We derive all equations of motion (EoMs) for the background and perturbations without any assumptions other than the linearized approximation for perturbations from the metric and scalar sectors of the model. Our analysis stands in stark contrast to most existing studies in the literature, which predominantly follow the slow-roll approximation. We provide a comparison between the slow-roll approximated values of the amplitude of the power spectrum of curvature perturbations and $n_s$, and their corresponding values estimated by directly solving the EoMs. This is of particular importance, given that both observables are measured with high precision.

A precise understanding of the post-inflationary reheating epoch is essential to match between the CMB scale, where the $\Lambda \mathrm{CDM}$ parameters such as $n_s$
are measured and, the scale where inflation took place. In this regard, it has been shown that the reheating
temperature is by far the most important parameter to account for the observed high $n_s$~\cite{Drees:2025ngb,Kallosh:2025ijd}.
Practically all studies conducted so far that aim to account for the high $n_s$ in the $R^2$-Higgs inflation (or $R^2$ inflation for that matter) 
treated the reheating temperature as a free parameter. However, a thorough analysis is still lacking to determine whether the parameter space yielding a high $n_s$ leads
to perturbative reheating or whether thermalization proceeds via preheating. E.g., it has been shown that the presence of $\xi_H \sim 10$, where $\xi_H$ is the 
non-minimal coupling between Higgs to Ricci scalar, may preheat the Universe via production of Higgs particles~\cite{Cado:2024von}.
In our work, we study the impact of preheating in the $R^3$ modified $R^2$-Higgs
inflation and its connection in matching the inflationary and the CMB scale. We construct gauge-invariant scalar perturbations considering all linear perturbations from scalar sector
as well as those from the metric and compute the EoMs utilizing covariant formalism. The EoMs are then solved to determine the corresponding preheating temperature to match the scales.
We show that the presence of $R^3$ term may lead to faster preheating via Higgs production than that of $R^2$-Higgs inflation without it. Our study illustrates that the $R^3$ term not 
only helps account for high $n_s$, it also impacts on the scale matching by inducing faster preheating.

The paper is organized as follows.
In Sec.\ref{sec:action}, we summarize the details of the model and derive the relevant EoMs, followed by details of the inflationary dynamics in Sec.\ref{sec:infdynamics}. The preheating dynamics, along with the estimation of the preheating temperature and its impact on matching the CMB reference scale, is discussed in Sec.\ref{sec:prehea}. We summarize our findings with an outlook in Sec.\ref{sec:disc}.

\section{The action}\label{sec:action}
The action of the $R^2$-Higgs inflation with the dimension six $R^3$ term in the Jordan frame is
\begin{align}
  S_J  &=  \int d^4 x \sqrt{-g_J} \bigg[ \frac{M_{\rm P}^{2}}{2} f(R_J, \Phi)
  -g_J^{\mu\nu}(\partial_\mu\Phi)^\dagger \partial_\nu\Phi - V(\Phi, \Phi^\dagger)
 \bigg],\label{eq:actionJ1}
\end{align}
where,
\begin{align}
f(R_J, \Phi) &= R_J  + \frac{\xi_R}{2 M_{\rm P}^2} R_J^2  + \frac{1}{3 M_{\rm P}^4 \xi_c} R_J^3 + \frac{2\xi_H}{M_{\rm P}^2} |\Phi|^2 R_J, \\\ V(\Phi, \Phi^\dagger) &= \lambda|\Phi|^4 \label{def:FR+CScouplings},
\end{align}
and, $M_{\rm P}=\sqrt{1/\left(8\pi G\right)}\approx 2.4\times 10^{18}~\text{GeV}$ is the reduced Planck mass with $G$ being Newton's gravitational constant
~\footnote{Note that, in references such as ~\cite{Ivanov:2021chn,Modak:2022gol,Addazi:2025qra} a different parametrization of $f(R)$ is followed $f(R)=\left(R + \frac{1}{6 M^2} R^2 + \frac{\delta_3}{36 M^4} R^3\right)$. Here parameters $M$ and $\delta_3$ is related to $\xi_R$ and $\xi_c$ as $M^2 = \frac{M_{\rm P}^2}{3 \xi_R}$ and $\delta_3= \frac{4}{ 3 \xi_R^2 \xi_c}$.}.
The chosen metric convention is $(-1,+1,+1,+1)$ and $\sqrt{-g_J}$ denotes the  determinant of the metric and $R_J$ is the space-time Ricci scalar.
The $\Phi$ is the Higgs field with hypercharge $+1$.
Note that here we ignored the mass term in the Higgs potential since it does not have any significant impact on the inflationary dynamics under consideration.

It is in general more convenient to study the dynamics of $R^2$-Higgs inflation in Einstein frame. To transform the  generic $f(R_J, \Phi)$ theory action in Eq.~\eqref{eq:actionJ1}
to the scalar-tensor theory we first introduce an auxiliary field $\Psi$ and perform a Legendre transformation.  The action is rewritten as
\begin{equation} \begin{aligned}
  S_J  =  \int d^4 x \sqrt{-g_J}  \bigg[&\frac{M_{P}^{2}}{2} \left(f(\Psi,  \Phi)
  + \frac{\partial f(\Psi,  \Phi)}{\partial \Psi} (R_J-\Psi)\right)
   -g_J^{\mu\nu}(\partial_\mu\Phi)^\dagger \partial_\nu\Phi - V(\Phi, \Phi^\dagger)\bigg]\label{eq:actionJ3}.
\end{aligned} \end{equation}
The Legendre transformation is well defined as long as $f(R,  \Phi)$ is convex~\cite{Ivanov:2021chn}. This leads to constrained relationship $\Psi > - \frac{1}{2} M_{\rm P}^2 \xi_c \xi_R$.
It is customary to introduce a physical degree of freedom
\begin{align}
\Theta = \frac{\partial f(\Psi,  \Phi)}{\partial \Psi},\label{eq:theta}
\end{align}
and rewrite the action in Eq.~\eqref{eq:actionJ3} as
\begin{equation} \begin{aligned}
  S_J  =  \int d^4 x \sqrt{-g_J} \bigg[& \frac{M_{P}^{2}}{2} \Theta R_J - U(\Theta,   \Phi)
  -g_J^{\mu\nu}(\partial_\mu\Phi)^\dagger \partial_\nu\Phi - V(\Phi, \Phi^\dagger)  \bigg].\label{eq:actionJ4}
\end{aligned} \end{equation}
Note that Eq.~\eqref{eq:theta} has two solutions and we have chosen 
\begin{align}
\Psi = - \frac{1}{2} M_{\rm P}^2 \xi_c \bigg(\xi_R- \zeta(\Theta, \Phi)\bigg)
\end{align}
where,
\begin{align}
 \zeta(\Theta, \Phi)=\sqrt{\xi_R^2 + \frac{4}{\xi_c}  \bigg(\Theta -1 - \frac{2 \xi_H \left(\Phi^\dagger \Phi\right)}{M_{\rm P}^2}\bigg)}.
\end{align}
The convexity condition here is satisfied during inflation if $\xi_c > 0$ for $\xi_R \gg \xi_H$. 
The potential $U(\Theta,   \Phi)$ with two degrees of freedom $\Theta$ and $\Phi$ takes the form
\begin{align}
U(\Theta,   \Phi) &= \frac{M_{\rm P}^2}{2}\left[\Psi \Theta -  f(\Psi,  \Phi)\right] = \frac{M_{\rm P}^4 \xi_c^2 }{48 }\bigg[\bigg(\xi_R- \zeta(\Theta, \Phi)\bigg) \bigg(\xi_R^2+ \xi_R \ \zeta(\Theta, \Phi) -2 \zeta(\Theta, \Phi)^2\bigg) \bigg].
\end{align}
For the other solution of Eq.~\eqref{eq:theta}, we remark that convexity is possible for $\xi_c < 0$, however 
negative $\xi_c$ may lead to unbounded $U(\Theta,   \Phi)$ from below.
After performing the Weyl rescaling of the metric $g^{\mu\nu}_J = \Theta \ g^{\mu\nu}_E$, we get the Einstein frame action
\begin{equation} \begin{aligned}
S_E  = \int d^4 x \sqrt{-g_E}\bigg[ & \frac{M_{\rm P}^2}{2} R_E - \frac{3 M_{\rm P}^2}{4} g_E^{\mu \nu} \partial_\mu (\ln\Theta) \partial_\nu(\ln\Theta)
-\frac{1}{2 \Theta} g_E^{\mu\nu}(\partial_\mu\Phi)^\dagger \partial_\nu\Phi - V_E \bigg]\label{eq:actionE1}
\end{aligned} \end{equation}
with 
\begin{align}
V_E &= \frac{1}{\Theta^2}\left[V(\Phi, \Phi^\dagger) +U(\Theta, \Phi)\right],\\
R_J &= \Theta \left[R_E +3 \Box_E  \Theta- \frac{3}{2} g_{E}^{\mu\nu} \partial_\mu (\ln\Theta) \partial_\nu(\ln\Theta) \right].
\end{align}
We have ignored the surface term $\Box_E = g_{E}^{\mu\nu} \partial_\mu \partial_\nu$ in the action $S_E$. With a field redefinition
\begin{align}
\phi = M_{\rm P} \sqrt{\frac{3}{2}} \ln\Theta,
\end{align}
and decomposing the Higgs field  in the Unitary gauge
\begin{align}
\Phi =
\frac{1}{\sqrt{2}} \begin{pmatrix}
  0 \\
  h  \\
\end{pmatrix} \label{eq:higgsuni},
\end{align}
the action Eq.~\eqref{eq:actionE1} takes the form
\begin{align}
S_E  &= \int d^4 x \sqrt{-g_E}\bigg[\frac{M_{\rm P}^2}{2} R_E - \frac{1}{2} G_{IJ} g_E^{\mu \nu} D_\mu \phi^I D_\nu \phi^J -
V_E(\phi^I) \bigg)\label{eq:actionfinal},
\end{align}
where,
\begin{align}
&V_E(\phi^I) = e^{-2\sqrt{\frac{2}{3}}\frac{\phi}{M_{\rm P}}}\Biggl[\frac{\lambda}{4} h^4 + \Biggl\{ \frac{M_{\rm P}^4 \xi_c^2 }{48 }\bigg[\bigg(\xi_R- \tilde{\zeta}(\phi,h)\bigg) \bigg(\xi_R^2+ \xi_R\tilde{\zeta}(\phi,h) -2 \tilde{\zeta}(\phi,h)^2\bigg) \bigg]\Biggr\}\Biggr],
\label{eq:pot}\\
&\qquad \qquad \mbox{with}~~
\tilde{\zeta}(\phi,h)=\sqrt{\xi_R^2 + \frac{4}{\xi_c}  \bigg(\Theta -1 - \frac{\xi_H h^2}{M_{\rm P}^2}\bigg)}.\nn
\end{align}
The $G_{IJ}$ is the  $2\times 2$ metric for the field space manifold $\phi^I \in \{\phi, h \}$ which has only two nonzero
elements $G_{\phi\phi} = 1$ and $G_{ h h} = e^{-\sqrt{\frac{2}{3}}\frac{\phi}{M_{\rm P}}}$.

\section{Inflationary dynamics}
\label{sec:infdynamics}

\subsection{Background and perturbation}
In order to study the inflationary dynamics, we first need to find the equation of motion (EoM)
of the scalar fields $\phi^I \in \{\phi, h \}$, which is obtained from varying the action Eq.~\eqref{eq:actionfinal}
\begin{equation} \begin{aligned}
&\Box \phi^K + \Gamma^{K}_{\ IJ} \ g_E^{\alpha \nu} D_\alpha  \phi^I D_\nu \phi^J - G^{KM} V_{E,M}
= 0\label{eom:scalar},
\end{aligned} \end{equation}
where $\Gamma^{K}_{\ IJ}$ are the Christoffel symbols for the field-space metric.
Due to the multifield nature of the action and presence non-canonical kinetic terms, we closely follow the
covariant formalism as detailed in Ref.~\cite{Gong:2011uw,Sfakianakis:2018lzf,Kaiser:2012ak} at the linear order in perturbation.
The fields $\phi^I(x^\mu)$ are decomposed into homogeneous background part (${\varphi}^I $) and perturbation ($\delta\phi^I$) as
\begin{align}
\phi^I(x^\mu) = \varphi^I(t) + \delta\phi^I(x^\mu)\label{fieldexpan},
\end{align}
where $t$ is the cosmic time. We denote background fields $\varphi^I(t) = \{\varphi(t),h_0(t)\}$, i.e., the $\{\varphi(t)$ and
$h_0(t)$ are the background fields  for $\phi(x^\mu)$ and $\phi^I(x^\mu)$ respectively.
The linear order perturbed Friedmann-Robertson-Walker (FRW) metric is given as~\cite{Kodama:1984ziu,Mukhanov:1990me,Malik:2008im}
\begin{align}
ds^2 &= -(1+2\mathcal{A}) dt^2 +2 a(t) (\partial_i \mathcal{B}) dx^i dt +
a(t)^2 \left[(1-2\psi) \delta_{ij}+ 2 \partial_i \partial_j \mathcal{E}\right] dx^i dx^j,\label{eq:frwmetric}
\end{align}
where $\mathcal{A}, \mathcal{B}, \psi$ and $\mathcal{E}$ are the metric scalar perturbations and $a(t)$
is scale factor. Throughout this work we consider the longitudinal gauge where the scalar perturbations $\mathcal{E}$ and $ \mathcal{B}$ vanish.

At the linear order of perturbation, one can simply find the EoMs of the background fields from Eq.~\eqref{eom:scalar}
\begin{align}
&\mathcal{D}_t \dot{\varphi}^I + 3 H\dot{\varphi}^I + G^{\phi J} V_{E,J}(\varphi^I)= 0\label{eq:bkg_inf},
\end{align}
where $\mathcal{D}_t$ and  $\mathcal{D}_J$ field space covariant derivatives~\cite{Gong:2011uw,Sfakianakis:2018lzf,Kaiser:2012ak}
\begin{align}
\mathcal{D}_t A^I &  = \dot{A}^I  + \Gamma^I_{\; JK} \dot{\varphi}^J A^K,\\
 \mathcal{D}_J A^I & = \partial_J A^I + \Gamma^I_{\; JK} A^K.
\end{align}
The Hubble parameter is defined as
\begin{align}
H^2 &= \left(\frac{\dot{a}}{a}\right)^2 = \frac{1}{3 M_{\rm P}^2} \bigg(\frac{1}{2} G_{IJ} \dot{\varphi}^I \dot{\varphi}^J + V_0(\varphi^I)\bigg),\label{hubble1}\\
\dot{H} &= -\frac{1}{2 M_{\rm P}^2} \bigg(G_{IJ} \dot{\varphi}^I \dot{\varphi}^J\bigg).\label{hubble2}
\end{align}
The slow-roll parameter $\epsilon$ and the number of $e$-foldings $\mathcal{N}$ are expressed as
\begin{align}
\epsilon = -\frac{\dot{H}}{H^2},
~~~\mathcal{N}(t) \equiv \ln \frac{a(t)}{a(t_{\rm{end}})},\label{epsiefol}
\end{align}
where $t_{\rm{end}}$ is the corresponding cosmic time when the inflation ends and  $a(t_{\rm{end}})$ is the scale factor at the end of inflation.
We denote the end of inflation when $\epsilon (t=t_{\rm{end}}) = 1$.
In what follows we shall use the $t$ and $\mathcal{N}$ interchangeably.
The energy density associated with background fields is
\begin{align}
&\rho_{\mathrm{inf}} = \frac{1}{2} G_{IJ} \dot{\varphi}^I \dot{\varphi}^J + V_0(\varphi^I),\label{eq:infenergy}
\end{align}
where $G_{IJ}$ is evaluated at the background field order.

\begin{figure}[h]
\begin{center}
\includegraphics[width=.32 \textwidth]{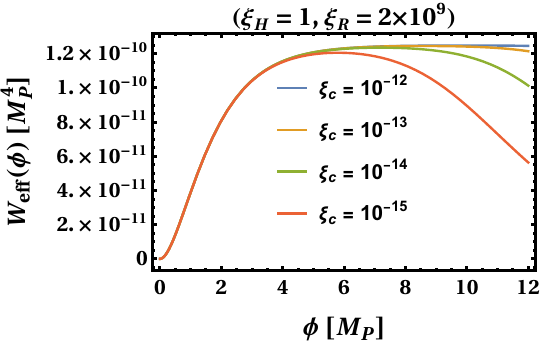}
\includegraphics[width=.32 \textwidth]{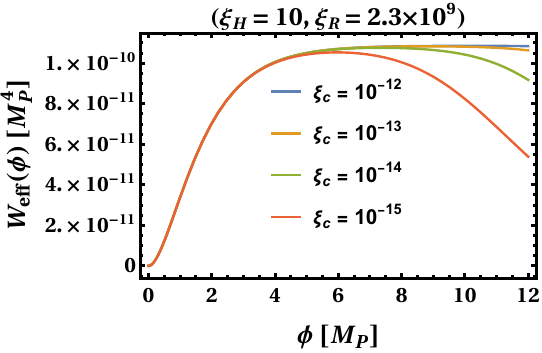}
\includegraphics[width=.32 \textwidth]{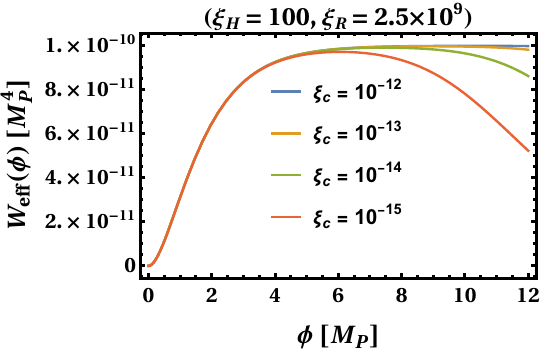}
\end{center}
\caption{The effective potential $W_{\rm{eff}}(\phi)$ vs $\phi$ plots for ($\xi_H = 1$, $\xi_R = 2\times 10^9$),  ($\xi_H = 10$,  $\xi_R = 2.3\times 10^9$) and ($\xi_H = 100$,  $\xi_R = 2.5\times 10^9$) for different values of $\xi_c$.}
\label{plot:potsing}
\end{figure}

Before moving to the perturbation dynamics let us first understand the impact of the $\xi_c$. Our primary aim is to
investigate whether the single field-like regime of $R^2$-Higgs inflation could still be a viable parameter space to account high $n_s$
after adding the $R^3$ term. The single field like regime can be obtained by so called valley approximation by solving
$\frac{\partial V_E}{\partial h} = 0$ for $h$  in terms of $\phi$ and, inserting the solution back to $V_E$~\cite{Lee:2023wdm}.
We denote this effective single field potential as $W_{\rm eff}(\phi)$. In the valley approximation,
for suitable values of $\xi_R$, $\xi_H$ and $\xi_c$, the $W_{\rm eff}(\phi)$ acts as effective inflaton potential
and $\phi$ plays the role of inflaton. In Fig.~\ref{plot:potsing}, we plot $W_{\rm eff}(\phi)$
for different $\xi_H$, $\xi_R$ and $\xi_c$.  The parameter $\lambda$ is kept
at a fixed value $10^{-2}$ throughout this paper for simplicity. It is clear that if $\xi_c$ becomes large the potential
become more flat i.e. akin to the case of pure $R^2$-Higgs inflation. For smaller $\xi_c$, the $W_{\rm eff}(\phi)$
does not remain asymptotically flat for large $\phi$. This is expected, since in
the $R^3$ term $\xi_c$ sits in the denominator (see Eq.~\eqref{def:FR+CScouplings}), hence a smaller $\xi_c$ impacts
more in changing the shape of $W_{\rm eff}(\phi)$. In the following we shall see that
$\xi_c\sim 10^{-13}$--$10^{-14}$ is sufficient to induce large $n_s$ to match the reported value of CMB+BAO.

\begin{table}[h]
\begin{tabular}{|c |c| c| c| c | c | c| c | c |c| c| c| c}
    \hline
	BP                  & $\xi_R$             &  $\xi_H$   & $\xi_c$                 &  $\varphi(t_{\text{in}})$ [$M_{\rm P}$]  & $h_0(t_{\text{in}})$  [$M_{\rm P}$] \\
   \hline
        $a$             & $2.12\times 10^9$   &  $1.5$     & $7\times 10^{-14}$      &  5.32                                    & $9.8\times10^{-5}$ \\
        $b$             & $2.3\times 10^9$    &  $10$      & $8\times 10^{-15}$      &  5.35                                    & $3\times10^{-7}$ \\
	\hline
	\end{tabular}
	\caption{Two benchmark points for our analysis. Scales are given in units of the Planck mass~$M_{\rm P}$. See text for details.}
	\label{parmeterchoices}
\end{table}

\begin{figure}[h]
\begin{center}
\includegraphics[width=.30 \textwidth]{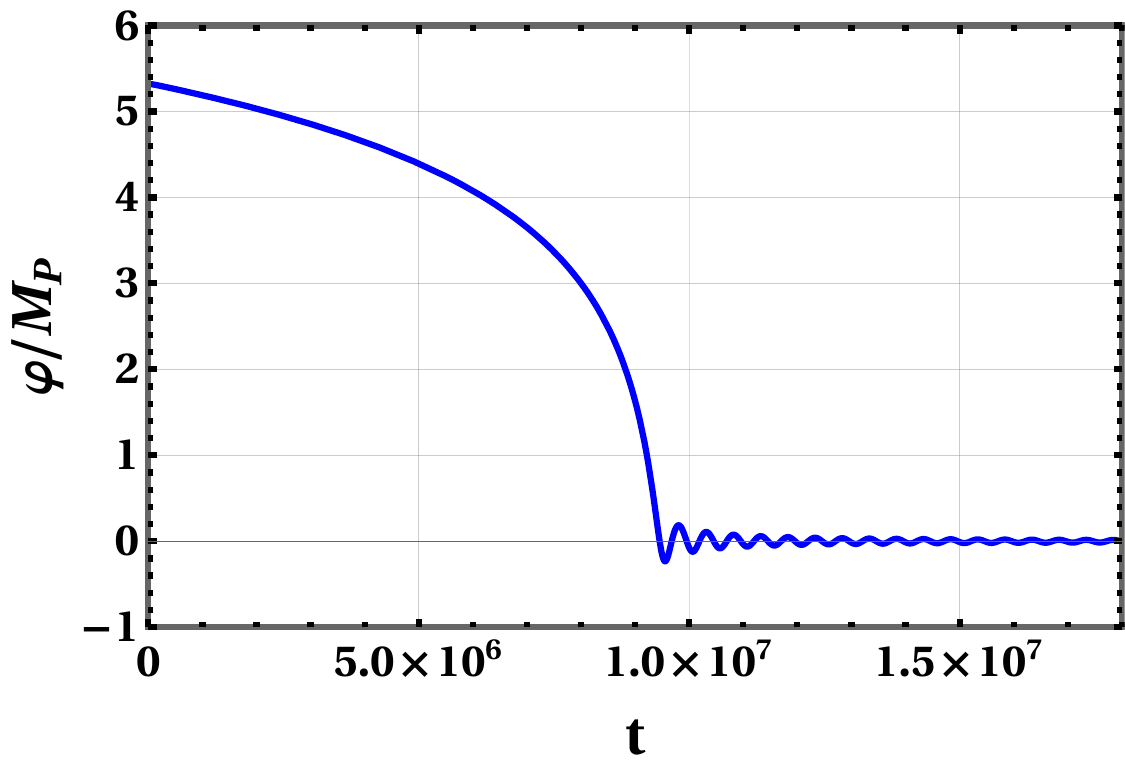}
\includegraphics[width=.33 \textwidth]{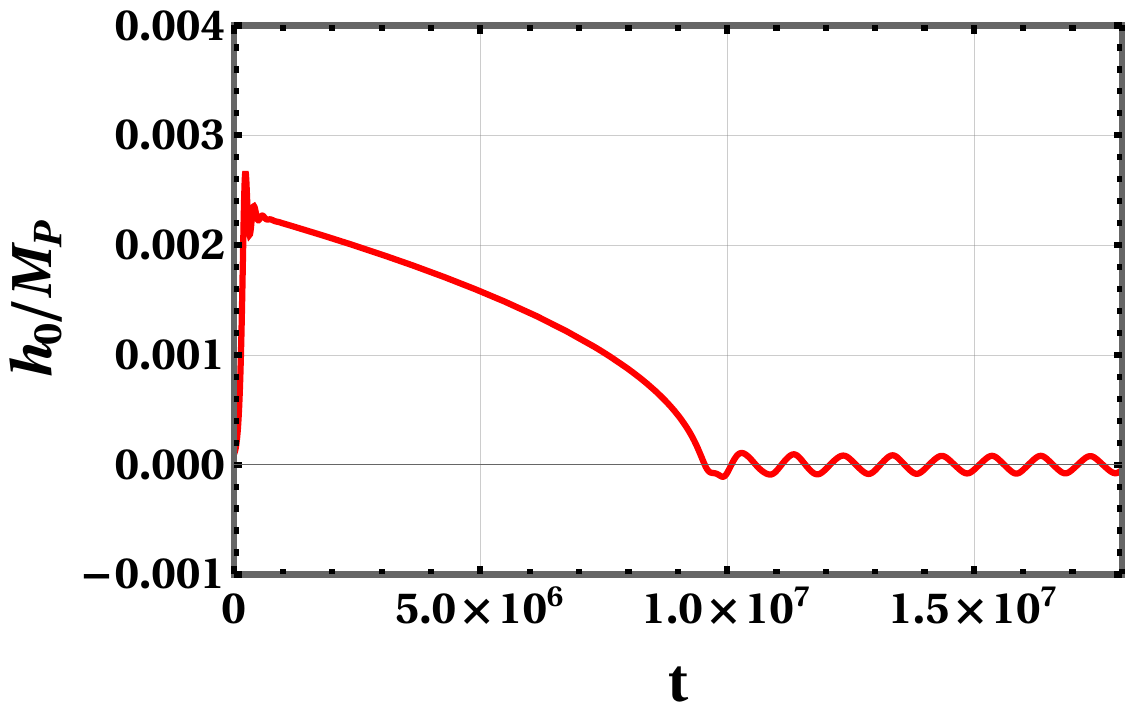}
\includegraphics[width=.35 \textwidth]{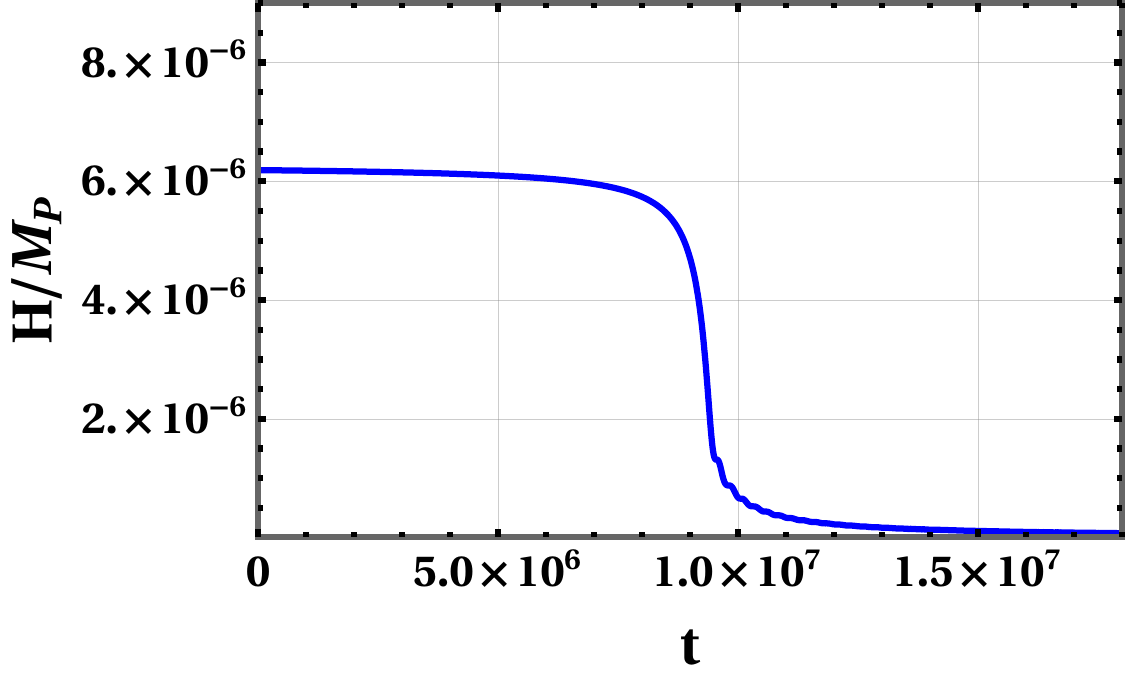}
\end{center}
\caption{The evolution of $\varphi$, $h_0$ and $H$ with respect to cosmological time $t$ for BP$a$.}
\label{plot:bkgevo}
\end{figure}

We utilize the valley approximation as a guideline to identify
a couple of single-field like benchmark points (BPs) as given in Table~\ref{parmeterchoices}.
In what follows, we take these BPs as reference and solve the coupled EoMs in Eq.~\eqref{eq:bkg_inf} directly
with the full potential $V_E(\phi^I)$. The initial time
for the numerical analysis is set to $t_{\text{in}}=0$. Note that Eq.~\eqref{hubble1} is solved simultaneously
taking $\ln(a)$ as a variable with initial condition $\ln(a(t_{\text{in}}))=0$. For consistency, we have checked the
$\dot{H}$ estimated from Eq.~\eqref{hubble2} matches with the one taking time derivative of Eq.~\eqref{hubble1}
within second decimal place. We remark here that the exact solution of the EoMs is required, rather than the
slow-roll approximation, due to the high precision of current measurements of parameters such as $n_s$.
In Fig~\ref{plot:bkgevo}, we plot the field evolution and Hubble parameter for BP$a$ for illustration.
Note that we solved the Eq.~\eqref{eq:bkg_inf} numerically with initial
conditions $\varphi(t_{\text{in}})$ as summarized in Table~\ref{parmeterchoices} and $\dot{\varphi}(t_{\text{in}})=0$.

\subsection{Dynamics of the perturbations}
The cosmological perturbations generated from the field fluctuations $\delta\phi^I(x^\mu)$ can account for the observed nearly scale invariant
adiabatic curvature perturbation. In general $\delta\phi^I(x^\mu)$
are gauge-dependent quantities but one can conveniently construct the covariant field fluctuations $\mathcal{Q}^I(x^\mu)$ which relates the $\phi^I(x^\mu)$
to corresponding $\varphi^I(t)$ via a unique geodesic in the field-space where
$\delta\phi^I$ is expressed as~\cite{Gong:2011uw,Elliston:2012ab}
\begin{align}
\delta\phi^I &= \mathcal{Q}^I -\frac{1}{2} \Gamma^I_{\ JK} \mathcal{Q}^K \mathcal{Q}^J+\frac{1}{3!} \big(\Gamma^I_{\ MN} \Gamma^N_{\ JK}-\Gamma^I_{\ JK,M}\big)  \mathcal{Q}^K \mathcal{Q}^J  \mathcal{Q}^M+\dots~.
\end{align}
At the linear order $\mathcal{Q}^I = \delta\phi^I$ and the gauge-independent Mukhanov-Sasaki variables
are~\cite{Sasaki:1986hm,Mukhanov:1988jd,Mukhanov:1990me}
\begin{align}
Q^I = \mathcal{Q}^I + \frac{\dot{\varphi}^I}{H}\psi = \delta\phi^I+ \frac{\dot{\varphi^I}}{H}\psi.\label{mukh-sasaki}
\end{align}
Note that is $Q^I$s are doubly covariant with respect both the space-time and field-space transformations.
In the field-space manifold $\dot{\varphi}^I$ and $Q^I$ transform like vectors.
We may now insert Eq.~\eqref{mukh-sasaki} and Eq.~\eqref{eq:frwmetric} into Eq.~\eqref{eom:scalar} to find
EoMs for $Q^I$s at linear order as
\begin{align}
\mathcal{D}_t^2 Q^I &+ 3 H \mathcal{D}_t Q^I -\frac{\nabla^2}{a^2} Q^I + \mathcal{M}^I_{\ \ J} Q^J = 0,\label{eqpertur:phi}
\end{align}
with
\begin{align}
\mathcal{M}^{I}_{\ L} = G^{IJ} (\mathcal{D}_L\mathcal{D}_J V_E)- \mathcal{R}^I_{\ JKL} \dot{\varphi}^J \dot{\varphi}^K
- \frac{1}{M_{\rm P}^2 a^3} \mathcal{D}_t \left(\frac{a^3}{H}\dot{\varphi}^I \dot{\varphi}_L\right),\label{eq:massterm}
\end{align}
where the $\mathcal{R}^I_{\ JKL}$ is the field-space Riemann tensor
which is evaluated, along with $\mathcal{M}^{I}_{\ L}$, at the background order.

The two independent perturbations $Q^\phi$ and $Q^h$ can be decomposed into adiabatic and isocurvature perturbations, however, we first need
two unit vectors $\hat{\sigma}^I$ and  $\hat{\omega}^I$. The former is defined as
\begin{align}
 \hat{\sigma}^I = \frac{\dot{\varphi}^I}{\dot\sigma},
\label{eq:sigmadotI}
\end{align}
with $\dot{\sigma} = \sqrt{G_{IJ} \dot{\varphi}^I \dot{\varphi}^J}$, while the latter  is defined as
\begin{align}
\hat{\omega}^I =\frac{\omega^I}{\omega},
\end{align}
where $\omega^I$ is called ``turning vector'' defined as $\omega^I =  \mathcal{D}_t  \hat{\sigma}^I$ and the magnitude
$\omega= |\omega^I|=\sqrt{G_{IJ} \omega^I\omega^J}$.
It should be noted that
$\omega_I\hat{\sigma}^I=0$.
We can now decompose the curvature perturbations and isocurvature perturbations as
\begin{align}
&Q_\sigma = \hat{\sigma}_I Q^I, \hspace{1cm}  Q_s = \hat{\omega}_I Q^I \label{eq:Q},
\end{align}
The gauge invariant curvature (adiabatic) and isocurvature perturbations are
\begin{align}
\mathcal{R} = \frac{ H}{\dot{\sigma}} Q_\sigma,\hspace{1cm}
\mathcal{S} = \frac{ H}{\dot{\sigma}} Q_s\label{eq:curventropy}.
\end{align}

The dimensionless power spectra for the adiabatic and entropy perturbations are~\cite{Mukhanov:1990me,Bassett:2005xm,Malik:2008im}
\begin{align}
&\mathcal{P}_{\mathcal{R}}(t;k)= \frac{k^3}{2\pi^2}|\mathcal{R}|^2\label{eq:powadia},\\
&\mathcal{P}_{\mathcal{S}}(t;k)= \frac{k^3}{2\pi^2}|\mathcal{S}|^2\label{eq:powentrop}.
\end{align}
One can now readily evaluate the power spectrum from Eq.~\eqref{eq:powadia} and Eq.~\eqref{eq:powentrop}
for a given Fourier mode $k$. The power spectrum for curvature perturbation freeze after it exits horizon.
Therefore we simply evaluate $\mathcal{P}_{\mathcal{R}}$ for different modes at the end of inflation.
The power spectrum of the isocurvature mode $\mathcal{P}_{\mathcal{S}}(t;k)$ on the other hand may change in the superhorizon scales.
This is primarily due to the small but non-vanishing off-diagonal elements of $M^I_{~J}$ which induces
mild power transfer from adiabatic to isocurvature mode. As we expect $\mathcal{P}_{\mathcal{R}}$ to be  orders of magnitude larger during inflation compared to
$\mathcal{P}_{\mathcal{S}}(t;k)$, a small power transfer from $\mathcal{P}_{\mathcal{R}}$ shall induce large change in $\mathcal{P}_{\mathcal{S}}(t;k)$.
Finally, the spectral index $n_{s}$ is given as
\begin{align}
n_{s} = 1 + \frac{d\ln\mathcal{P}_{\mathcal{R}}(k)}{d\ln k}\label{specin1}.
\end{align}

The measured values of inflationary parameters are~\cite{SPT-3G:2025bzu,BICEP:2021xfz}
\begin{align}
 & \log(10^{10}\mathcal{A}_{s*}) = 3.0574 \pm 0.0094\\
 & n_s^* = 0.9728 \pm 0.0027\\
 & r_* \lesssim 0.036~\rm{at~95\%~CL},
\end{align}
where $*$ denotes the respective parameters are measured at a reference
scale $k_{\rm ref} = 0.05~\rm{Mpc}^{-1}$ and $\mathcal{A}_{s*}$ is amplitude of the power spectrum of the curvature perturbation evaluated
at $k_{\rm ref}$. Here we have considered CMB-SPA+ DESI from Ref~\cite{SPT-3G:2025bzu} which includes all CMB measurements as well as BAO data from DESI.

\begin{figure}[h]
\begin{center}
\includegraphics[width=.36 \textwidth]{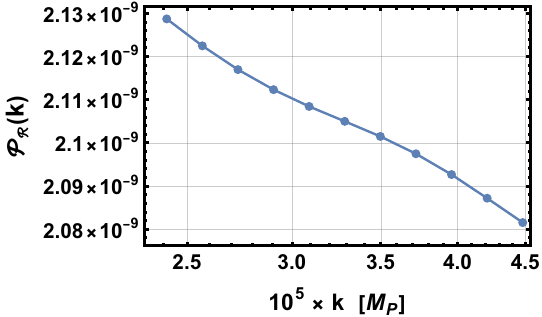}\hspace{1.5cm}
\includegraphics[width=.33 \textwidth]{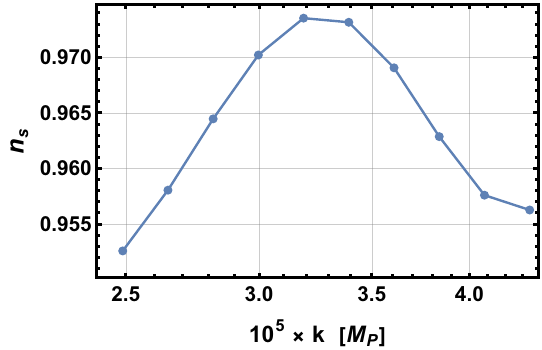}\\
\includegraphics[width=.38 \textwidth]{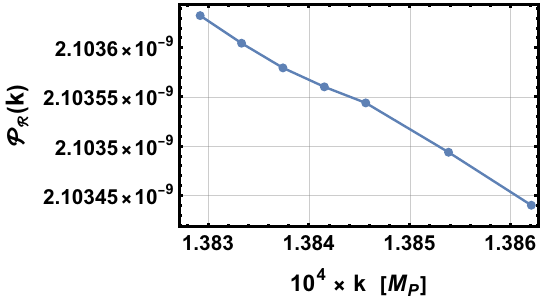}\hspace{1.9cm}
\includegraphics[width=.33 \textwidth]{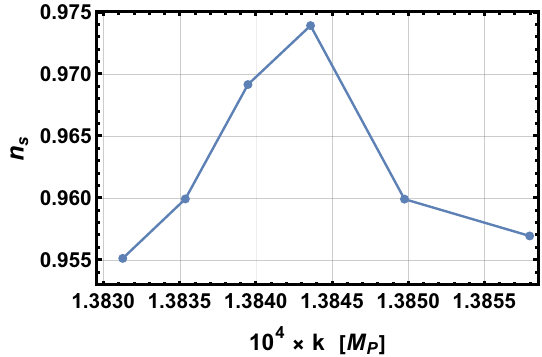}\\
\end{center}
\caption{The $\mathcal{P}_{\mathcal{R}}(k)$ and $n_s$ vs $k$ for both the BPs. Here the $n_s$ is evaluated directly utilizing Eq.~\eqref{specin1} and $\mathcal{P}_{\mathcal{R}}(k)$ via Eq.~\eqref{eq:powadia}.}
\label{plot:prns}
\end{figure}

\begin{figure}[h]
\begin{center}
\includegraphics[width=.35 \textwidth]{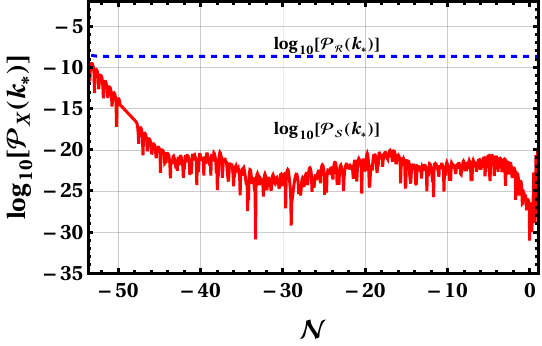}\hspace{1cm}
\includegraphics[width=.35 \textwidth]{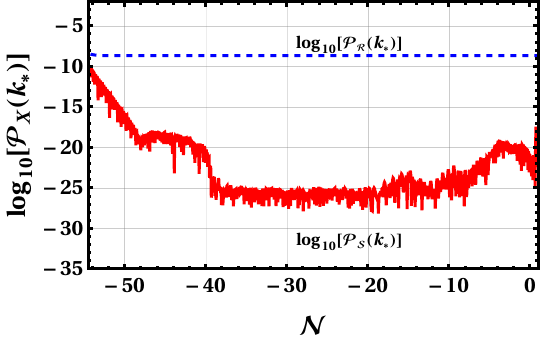}
\end{center}
\caption{The evolution of $\mathcal{P}_{\mathcal{R}}(k_*)$  and $\mathcal{P}_{\mathcal{S}}(k_*)$ after horizon exit for both the BPs.}
\label{plot:PRPS}
\end{figure}

In Fig.~\ref{plot:prns} we show the $\mathcal{P}_{\mathcal{R}}$ and $n_s$ as function of
Fourier modes $k$ evaluated from Eq.~\eqref{eq:powadia} and Eq.~\eqref{specin1}.
The reference mode $k_{\rm ref} = 0.05~\rm{Mpc}^{-1}$ today correspond to a mode $k_*=a(t_*) H(t_*)$ which exit the horizon at time $t_*$ before end of inflation
such that the corresponding power spectrum $\mathcal{P}_{\mathcal{R}}(t_*;k_*) = \mathcal{A}_{s*}$ and ${n_{s}}$ evaluated
at $k = k_*$ is equals to $n_s^*$. From Fig.~\ref{plot:prns}, we find that
 the $k_* = 3.4624\times 10^{-5}~(1.3843\times 10^{-4})~M_{\rm{P}}$,
 for BP$a$ (BP$b$), which correspond to $t_*= 2.784\times 10^5~ (5.3037\times 10^5)$ and, $\mathcal{P}_{\mathcal{R}}(k_*) = 2.102\times10^{-9} (2.104\times10^{-9})$
and $n_s^*=0.9714~(0.9733)$. The tensor-to scalar ratio is estimated simply utilizing single-field approximation $r\approx  16 \epsilon(t_*)$. It is justified since the
observable is not measured and only upper limit exist. We find $r_* \approx 3.75\times 10^{-3}~(3.45\times 10^{-3})$  for the respective BPs.

To generate Fig.~\ref{plot:prns}, we note that the EoMs in Eq.~\eqref{eqpertur:phi} are solved
for each Fourier modes $k$ using initial condition~\cite{Antusch:2015nla}
\begin{align}
Q^I(t_{\rm in})  \simeq \frac{H}{\sqrt{2 k^3}}\bigg(i+ \frac{k}{a H}\bigg)e^{i\frac{k}{a H}}\label{inifluc1},
\end{align}
with all relevant modes are initialized at $t=t_{\rm in}$ to ensure they are well within the horizon.
In Fig.~\ref{plot:PRPS}, we plot the evolution of $\mathcal{P}_{\mathcal{R}}(k_*)$  and $\mathcal{P}_{\mathcal{S}}(k_*)$ as function of cosmic time.
For both BPs, as expected, $\mathcal{P}_{\mathcal{S}}$ is orders of magnitude lower that of $\mathcal{P}_{\mathcal{R}}$ during inflation for the reference mode $k_*$.
Moreover, as explained above, $\mathcal{P}_{\mathcal{R}}(k_*)$ remains frozen but, $\mathcal{P}_{\mathcal{S}}(k_*)$ changes
during inflation, albeit its magnitude is still much smaller than the former.
The number of $e$-folding when the reference mode exit horizon before the end of inflation is $\Delta N_* = \mathcal{N}\bigr|_{t=t*}-\mathcal{N}_{t=t_{\rm end}}$, which is 53.66 and 54.51 for BPb and BPc respectively. At this point it is useful to compare the $\mathcal{A}_{s*}$ and $n_{s*}$ values obtained above to the ones from slow-roll approximation. In slow-roll approximation
\begin{align}
&\mathcal{A}_{s}(t) = \frac{V_E(t)}{24  \pi^2  M_{\rm P}^4 \ \epsilon(t) }\\
&n_{s}(t) \approx 1 -6 \epsilon(t) + 2 \eta(t) \label{specin2}.
\end{align}
Once evaluated at $t=t_*$ these expressions translate to $n_s^* = 0.9652$ (0.9746) and $\mathcal{A}_{s^*}=2.068\times10^{-9}$ ($2.0465\times10^{-9}$) for BP$a$ (BP$b$).
This emphasizes that the slow-roll approximation is perhaps not ideal given the high precision recent cosmological measurements.

Task now is to match the reference mode $k_*$ in unit of $M_{\rm P}$ to CMB reference mode $k_{\rm ref}/a_0 = 0.05~\rm{Mpc}^{-1}$. The $k_{\rm ref}$ is connected
to $k_*$ via the relationship
\begin{align}
k_{\rm ref} =  k_* = a(t_*) H(t_*)= \frac{a(t_*)}{a(t_{\rm end})} \frac{a(t_{\rm end})}{a(t_{\rm pre})} \frac{a(t_{\rm pre})}{a_0} a_0 H(t_*), \label{eq:kefrel}
\end{align}
where, $a(t_{\rm pre})$ is the corresponding scale factor when preheating is completed. Here we simply assumed that the radiation dominated era has immediately 
started after the end of preheating. The explicit details of the preheating epoch for the BP$a$ and BP$b$ is discussed shortly.
One can re-express Eq.~\eqref{eq:kefrel} as~\cite{He:2020ivk}
\begin{align}
\Delta N_* = \ln\left(\frac{a(t_{\rm end})}{a(t_*)}\right)&= \ln\left(\frac{ H(t_*)}{\left(k_{\rm ref}/a_0\right)}\right)+\ln\left(\frac{a(t_{\rm end})}{a(t_{\rm pre})}\right)
                                                                             +\ln\left(\frac{a(t_{\rm pre})}{a_0}\right) \nn\\
=& \ln\left(\frac{ H(t_*)}{\left(k_{\rm ref}/a_0\right)}\right)- \mathcal{N}_{\rm pre} +\ln\Biggl\{\frac{T_0}{T_{\rm pre}}\left(\frac{g_0}{g_{\rm{pre}}}\right)^{1/3}\Biggr\},\label{eq:nstar}
\end{align}
where $T_{\rm pre}$ and $T_0=2.7 K$  are the temperature at the end of preheating and today.  The number of relativistic degrees of freedom at the
end of preheating and today are $g_{\rm{pre}}=106.75$ and $g_0= 43/11$ respectively
and $\mathcal{N}_{\rm pre} = \ln\left(a(t_{\rm pre})/a(t_{\rm end})\right)$ is the number of $e$-folding elapsed from the end of inflation to the end of preheating.
Note here that while finding Eq.~\eqref{eq:nstar} we have assumed that the thermalization after is completed within one Hubble time as in Ref.~\cite{He:2020ivk}.
We may now express all dimensionfull quantities in $M_{\rm P}$ unit and match the right hand side of Eq.~\eqref{eq:nstar} i.e. $\Delta N_*$ to the left hand side.
While $\Delta N_*$ is already found above, parameters $\mathcal{N}_{\rm pre}$ and $T_{\rm pre}$ are yet to be estimated.
A detailed discussion on them is deferred to Sec.~\ref{sec:prehea}.

\section{Preheating}
\label{sec:prehea}
In order to find preheating we write the second order action for the fluctuations $Q^I$ (with $I=\{1,2\}$) as~\cite{Gong:2011uw,DeCross:2015uza,DeCross:2016cbs,Sfakianakis:2018lzf}
\begin{align}
S^{(2)}_{(Q)} =& \int d^3x \ dt \ a^3
\bigg[-\frac{1}{2}\overline{g}^{\mu\nu}_{E} G_{IJ} \mathcal{D}_\mu Q^I\mathcal{D}_\nu Q^J-\frac{1}{2}\mathcal{M}_{IJ} Q^I Q^J
\bigg],\label{action:quadfluc}
\end{align}
where $\overline{g}^{\mu\nu}_{E}\equiv(-1,a^2(t), a^2(t), a^2(t))$ is the spatially-flat unperturbed FLRW metric
and $\mathcal{M}_{IJ}$ and $G_{IJ}$ are computed at the background order. However, it is more convenient to quantize the fields for preheating in conformal time $\tau$ with invariant
$ds^2 = a^2(\tau) \eta_{\mu\nu} dx^\mu dx^\nu$. The second order action in conformal time with transformation $\partial_0 \to \partial_\tau / a$ and field rescaling $X^I(x^\mu) \equiv a(t) Q^I(x^\mu)$ is given as 
\begin{align}
S^{(2)}_{(X)} =& \int d^3x \ d\tau \Biggl[-\frac{1}{2} \eta^{\mu\nu} G_{IJ} (\mathcal{D}_\mu X^I)(\mathcal{D}_\nu X^J)-\frac{1}{2}\mathscr{M}_{IJ} X^I X^J\Biggr], 
\label{action:S_X2-inf}
\end{align}
with $\eta^{\mu\nu}=(-1,1, 1, 1)$.
\begin{align}
\mathscr{M}_{IJ} =a^2\bigg(\mathcal{M}_{IJ} -\frac{1}{6} G_{IJ}R_E\bigg),~~\mbox{with}~~R_E = \frac{6a''}{a^3},
\label{equ:mathcal-MIJ-def}
\end{align}
where we have used the shorthand notation $(')$ conformal time derivative.
The EoMs of the scalar field fluctuations $X^I(x^\mu)$ can now readily be derived from either from Eq.~\eqref{action:S_X2-inf} or from Eq.~\eqref{eqpertur:phi}
\begin{align}
\mathcal{D}_\tau^2 X^I &-\bigg[\nabla^2  - a^2 \left(\mathcal{M}^I_{\ \ I} -\frac{1}{6}R_E G^I_{\ \ I} \right)\bigg] X^I = 0,\label{eqpertur:X}
\end{align}
where we have used the diagonality of $M^I_{~~J}$. The energy momentum tensor for doubly covariant conformally rescaled field fluctuations $X^I$s is 
\begin{align}
T_{\mu\nu}^{(X)} =\ & G_{IJ} (\mathcal{D}_\mu X^I)(\mathcal{D}_\nu X^J)
+ \eta_{\mu\nu} \Biggl[-\frac{1}{2} \eta^{\alpha\beta} G_{IJ} (\mathcal{D}_\alpha X^I)(\mathcal{D}_\beta X^J)-\frac{1}{2}\mathscr{M}_{IJ} X^I X^J
 \Biggr].\label{eq:infHiggflucstress}
\end{align}

\begin{figure}[h]
\begin{center}
\includegraphics[width=.44 \textwidth]{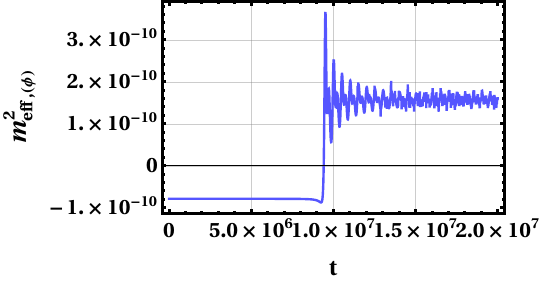}\hspace{1cm}
\includegraphics[width=.44 \textwidth]{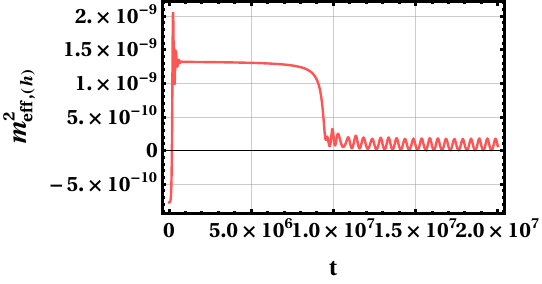}\\
\includegraphics[width=.44 \textwidth]{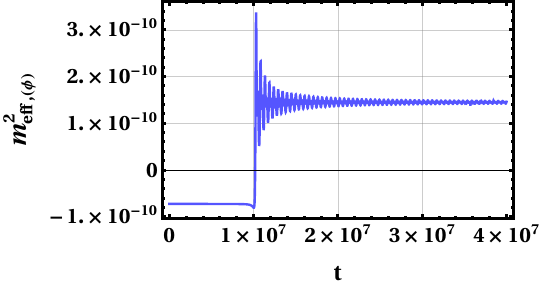}\hspace{1cm}
\includegraphics[width=.44 \textwidth]{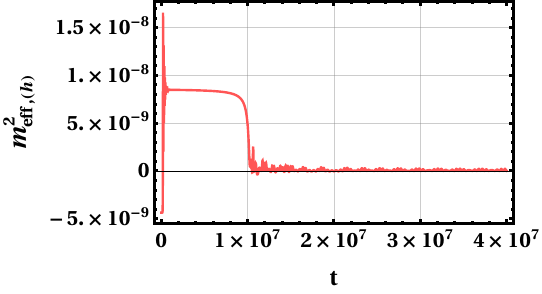}
\end{center}
\caption{The evolution of $m_{\mathrm{eff},(I)}^2(\tau)$ as function of $\mathcal{N}$. See text for details.}
\label{plot:meffplot}
\end{figure}

We transform the second order action in Eq.~(\ref{action:S_X2-inf}) to momentum space~\cite{DeCross:2015uza,DeCross:2016cbs,Sfakianakis:2018lzf,Cado:2024von,Cado:2023zbm}
\begin{align}
S_{(X)} =&  \int d\tau \, \mathcal{L}_{(X)} = \int \, d\tau \,\frac{d^3k}{(2\pi)^3}\, \ \bigg[\frac{1}{2} \left|\partial_\tau  \widetilde{X}^I\right|^2
- \frac{1}{2}\omega_{(I)}^2(\tau,k) \left|\widetilde{X}^I\right|^2 \bigg], \label{action:quanphih}
\end{align}
with
\begin{align}
\omega_{(I)}^2(\tau,k) = \left(k^2 + a^2 m_{\mathrm{eff},(I)}^2(\tau)  \right)~\mbox{and}
~m_{\mathrm{eff},(I)}^2(\tau)= \mathcal{M}^I_{~~I} - \frac{1}{6}R_E = \frac{1}{a^2}\; \mathscr{M}^I_{~~I} = m_{\mathrm{eff},(I)}^2= \sum_i m_{i,(I)}^2
\label{def:omega2}
\end{align}
where we have denoted
\begin{align}
m_{1,(I)}^2 =G^{(I)J} (\mathcal{D}_{(I)}\mathcal{D}_J V_E),~~
m_{2,(I)}^2 =- \mathcal{R}^{(I)}_{\ \ JK(I)} \dot{\varphi}^J \dot{\varphi}^K,~~
m_{3,(I)}^2 =- \frac{1}{M_{\rm P}^2 a^3} \mathcal{D}_t \left(\frac{a^3}{H}\dot{\varphi}^{(I)} \dot{\varphi}_{(I)}\right),~~
m_{4,(I)}^2 =-\frac{R_E}{6},  \label{def:effective-masses-decomposition}
\end{align}
and the field-space indices $(I)$ are not summed over. 
In Fig.~\ref{plot:meffplot} the $m_{\mathrm{eff},(I)}^2(\tau)$ as function of $\mathcal{N}$ is plotted for illustration. We see that in both the BPs
$m_{\mathrm{eff},(\phi)}^2(\tau)$ is tachyonic but becomes positive after the end of inflation. In contrast $m_{\mathrm{eff},(h)}^2(\tau)$ is positive
before end of inflation while oscillates around zero after end of inflation. We postpone a discussion regarding consequences of them for later part of this section.

The canonical momentum is defined as
\begin{align}
&\hat{\widetilde{\pi}}^{I}(\tau,\vb{k})=  \partial_\tau \hat{\widetilde{X}}^{I}(\tau,\vb{k}), ~\mbox{with}~\bigg[ \hat{\widetilde{X}}^{I}(\tau,\vb{k}),\hat{\widetilde{\pi}}^{J}(\tau,\vb{q})\bigg]= i (2\pi)^3\delta^{IJ}\delta^{(3)}(\vb{k}+\vb{q}) \label{quant:phih},
\end{align}
where we have elevated the classical field fluctuations $\widetilde{X}^{I}$ to their 
respective quantized versions $\hat{\widetilde{X}}^{I}$.
We now decompose the quantized fluctuations $\hat{\widetilde{X}}^\phi$ and
$\hat{\widetilde{X}}^h$ in momentum space as~\cite{DeCross:2015uza,Sfakianakis:2018lzf}
\begin{align}
&\hat{\widetilde{X}}^\phi =   \left[ \left(v_{1k}(\tau) e_1^\phi(\tau) \hat{a}_1(\vb{k}) + v_{2k}(\tau) e_2^\phi(\tau) \hat{a}_2(\vb{k})\right)
+ \left(v^*_{1k}(\tau) e_1^\phi(\tau) \hat{a}^\dagger_1(-\vb{k}) + v_{2k}^*(\tau) e_2^\phi(\tau) \hat{a}^\dagger_2(-\vb{k})\right)   \right],\label{eq:phiquant}\\
&\hat{\widetilde{X}}^h =  \left[ \left(y_{1k}(\tau) e_1^h(\tau) \hat{a}_1(\vb{k}) + y_{2k}(\tau) e_2^h(\tau) \hat{a}_2(\vb{k})\right)
+  \left(y^*_{1k}(\tau) e_1^h(\tau) \hat{a}^\dagger_1(-\vb{k}) + y^*_{2k}(\tau) e_2^h(\tau) \hat{a}^\dagger_2(-\vb{k})\right)  \right],\label{eq:hquant}
\end{align}
$\hat{a}_m(\vb{k})$ and $\hat{a}^\dagger_m(-\vb{k})$ (with $m\in{1,2}$) are annihilation and creation operators which follow
the  commutator relationships
\begin{align}
\left[\hat{a}_m(\vb{k}), \hat{a}_n(\vb{q})\right] = \left[\hat{a}^\dagger_m(\vb{k}), \hat{a}^\dagger_n(\vb{q})\right] = 0, \hspace{2 cm} 
\left[\hat{a}_m(\vb{k}), \hat{a}^\dagger_n(\vb{q})\right] = (2\pi)^3 \delta_{mn} \delta^{(3)}(\vb{k}-\vb{q}),
\end{align}
with
\begin{align}
 \hat{a}_m(\vb{k})  \ket{0} = 0,\hspace{3 cm} \bra{0} \hat{a}^\dagger_m(\vb{k}) = 0.
\end{align}
The field space vielbeins follow relationship
\begin{align}
\delta^{mn} e_m^I(\tau) e_n^J(\tau) = G^{IJ}(\tau),~\mbox{and}\mathcal{D}_\tau  e^m_J = 0,\label{eq:vielbein-conditon}
\end{align}
for all $m$ and $J$.

As our focus of interest is the single field like regime, the off-diagonal elements $\mathcal{M}^\phi_{~~h} \sim 0$ and $\mathcal{M}_{~~\phi}^h \sim 0$ and, hence
$e_2^\phi\sim 0$, $e_1^h \sim 0$~\cite{DeCross:2016cbs,Cado:2024von}. Therefore, Eq.~\eqref{eqpertur:X} 
becomes two decoupled source-free EoMs as
\begin{align}
&v_{1k}'' + \omega^2_{(\phi)}\,v_{1k} \simeq 0,\label{eq:inflatonfluc}\\
&y_{2k}'' + \omega^2_{(h)}\,y_{2k} \simeq 0,\label{eq:higgsfluc}
\end{align}
with $\omega^2_{(I)} $ given by Eq.~\eqref{def:omega2}.
We solve the EoMs Eq.~\eqref{eq:inflatonfluc} and Eq.~\eqref{eq:higgsfluc} utilizing the  Bunch-Davis (BD) initial condition
\begin{align}
\lim_{t \to - \infty}v_{1k}(k,t) = \lim_{t \to - \infty}y_{2k}(k,t) = \frac{ e^{-\frac{i k t}{a}}}{\sqrt{2 k}}, \hspace{1.cm}
\lim_{t \to - \infty}\dot{v}_{1k}(k,t)=\lim_{t \to - \infty}\dot{y}_{2k}(k,t) = - \frac{i}{a}  \sqrt{\frac{k}{2}} \; e^{-\frac{i k t}{a}}\label{eq:ini},
\end{align}
where the relevant modes under consideration are initialized about $\mathcal N\sim -3$ i.e. $3$ $e$-foldings before end of inflation
to ensure they are well within the horizon.

The combined vacuum averaged comoving energy densities for the inflaton and Higgs fluctuations is~\cite{Cado:2024von,Cado:2023zbm}
\begin{align}
\rho_{\phi h}
= \int \frac{d^3k}{(2\pi)^3} \ \left(\rho_k^{(\phi)}+ \rho_k^{(h)}\right), \label{eq:energydensity}
\end{align}
where $\rho_k^{(\phi)}$ and $\rho_k^{(h)}$ are the respective fluctuations per mode defined as
\begin{subequations} \begin{eqnarray}
\rho_k^{(\phi)} &=& \dfrac{1}{2} G_{\phi\phi}\left(|v'_{1k}|^2 +\omega^2_{(\phi)}   |v_{1k}|^2\right) e_1^\phi e_1^\phi = \dfrac{1}{2}\left(|v'_{1k}|^2 +\omega^2_{(\phi)}   |v_{1k}|^2\right) , \label{enphi_conf-simp}\\
 \rho_k^{(h)} &=& \dfrac{1}{2} G_{hh}\left(|y'_{2k}|^2 +\omega^2_{(h)}  |y_{2k}|^2 \right)e_2^h e_2^h = \dfrac{1}{2}\left(|y'_{2k}|^2 +\omega^2_{(h)}  |y_{2k}|^2 \right). \label{enh_conf-simp}
\end{eqnarray}  \end{subequations}
The physical energy densities of the fluctuation of $\phi$ is
\begin{align}
\rho_{(\phi)}=  \frac{1}{a^4}\int \frac{d^3k}{(2\pi)^3} \rho_k^{(\phi)}
&= \dfrac{1}{a^2} \displaystyle\int \dfrac{k^2}{4\pi^2} dk \left[|\dot{v}_{1k}|^2 + \left(\dfrac{k^2}{a^2}+ \left| m^2_{\mathrm{eff},(\phi)}(t) \right| \right) |v_{1k}|^2\right], \label{eq:energy_denphi}
\end{align}
and a similar expression can be found for $h$
The vacuum subtracted quantum energy densities are given as~\cite{Cado:2024von} 
 \begin{align}
\rho^q_{(I)} &= \rho_{(I)}- \rho_{(I)}^{\rm BD},~\mbox{with}~~\rho_{(I)}^{\rm BD}   = \frac{1}{a^4}\int dk \;  \frac{k^3}{4\pi^2},
 \label{eq:quantumphi}
\end{align}  
where,  $\rho_{(I)}^{\rm BD}$ is the associated BD vacuum energy densities for respective field.

\begin{figure}[h]
\begin{center}
\includegraphics[width=.37 \textwidth]{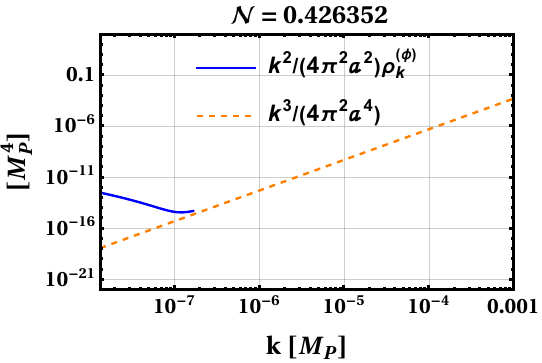}\hspace{1cm}
\includegraphics[width=.37 \textwidth]{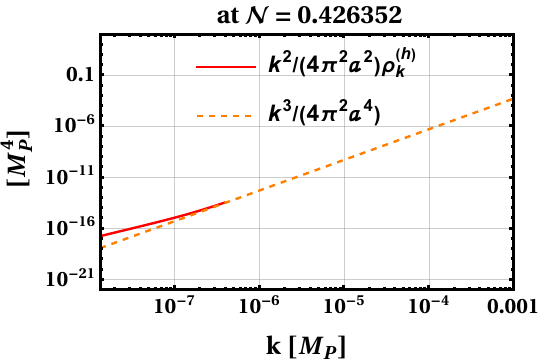}\\
\includegraphics[width=.37 \textwidth]{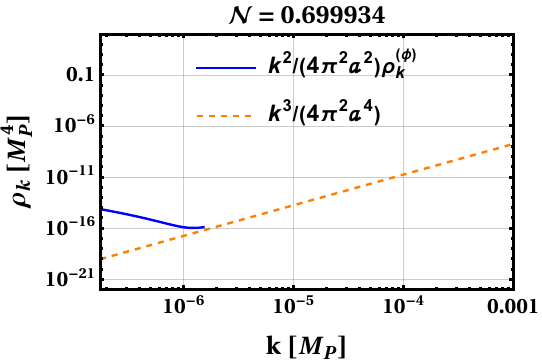}\hspace{1cm}
\includegraphics[width=.37 \textwidth]{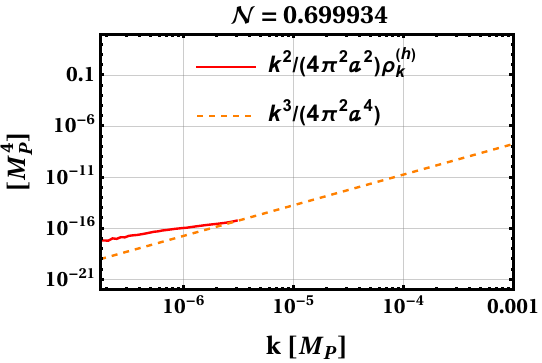}\\
\end{center}
\caption{The spectrum for inflaton  fluctuations ($k^2/\left(4\pi^2 a^2\right)\rho_k^{(\phi)}$) 
and the BD vacuum ($k^3/\left(4\pi^2 a^4\right)$) are plotted in the left panels for both
the BPs in solid blue and dotted orange respectively.
In right panels the corresponding spectrum of Higgs ($k^2/\left(4\pi^2 a^2\right)\rho_k^{(h)}$) fluctuations are plotted in solid
against the BD vacuum in dotted orange. Note that, these spectra are essentially the integrand of Eq.~\eqref{eq:energy_denphi} and Eq.~\eqref{eq:quantumphi} for the respective fluctuations and BD vacuum.}
\label{plot:spectr}
\end{figure}
\begin{figure}[h]
\begin{center}
\includegraphics[width=.35 \textwidth]{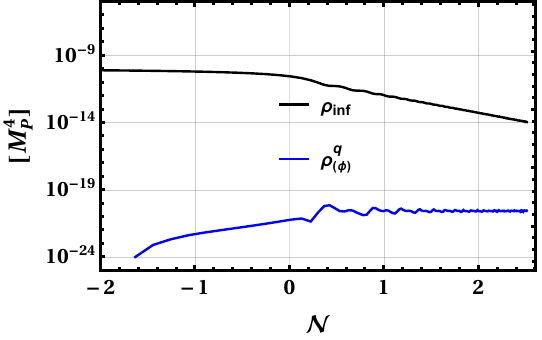}\hspace{1cm}
\includegraphics[width=.35 \textwidth]{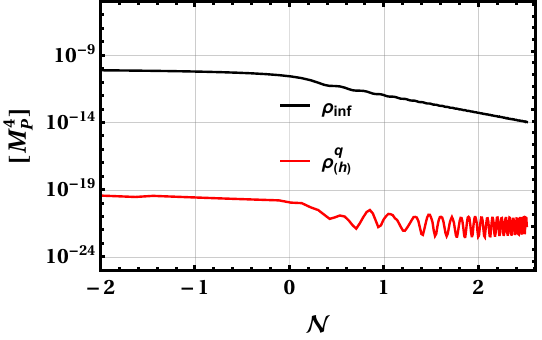}\\
\includegraphics[width=.35 \textwidth]{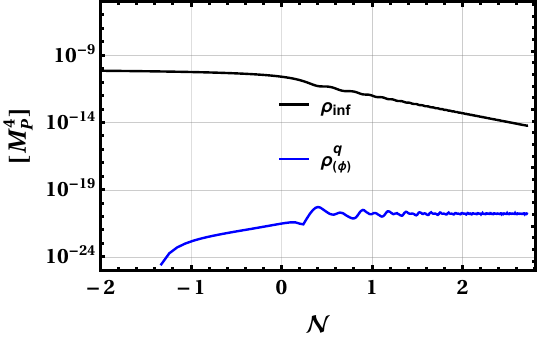}\hspace{1cm}
\includegraphics[width=.35 \textwidth]{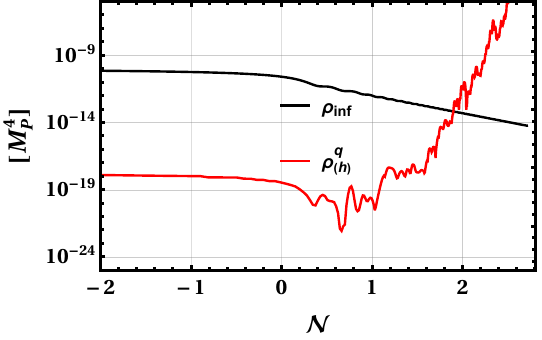}
\end{center}
\caption{The vacuum subtracted quantum energy densities $\rho^q_{(\phi)}$ and $\rho^q_{(h)}$ for BP$a$ (upper panel) and BP$b$ (lower panel) respectively along with
$\rho_{\rm inf}$ for comparison.}
\label{plot:energy}
\end{figure}

The momentum upper limit in Eq.~\eqref{eq:energy_denphi} and Eq.~\eqref{eq:quantumphi} are obtained by finding the mode for which the
relative error between $\rho_k$ and $\rho_k^{\rm BD}$ is about 10\% with $\rho_k >\rho_k^{\rm BD}$ in each time
step. For illustration, we have plotted the spectrum in Fig.~\ref{plot:spectr} for different $\mathcal{N}$.
The momentum upper limit is identified as the $k$ value where the blue or red curves stop respectively
slightly above the orange dotted BD spectrum (within 10\% of each other with $\rho_k >\rho_k^{\rm BD}$) for specific $\mathcal{N}$~\cite{Cado:2024von}.
Modes larger than this $k$ do not leave the vacuum in the respective time steps.
The lower limit on the other hand is found by considering all
sub-horizon modes at a particular time step, since only these sub-horizon modes take part in thermalization process
in our single-filed like scenario. In practice, we follow an adaptive numerical framework where we consider all sub-horizon modes
at a particular time while the upper limit is obtained from the above mentioned method. We redirect readers
to Ref.~\cite{Cado:2024von} for details of this numerical approach.

We plot $\rho^q_{(\phi)}$ (blue), $\rho^q_{(h)}$ (red), and $\rho_{\rm inf}$ (black) in Fig.~\ref{plot:energy} for illustration.
It is clear that $\rho^q_{(\phi)}$ is orders of magnitude lower than $\rho_{\rm inf}$ for both BP$a$ and BP$b$. This is primarily because
$m_{\mathrm{eff,(\phi)}}^2 < 0$; as a result, $\rho^q_{(\phi)}$ receives a tachyonic (exponential) amplification before the end of inflation
for both BPs. However, after the end of inflation, $m_{\mathrm{eff,(\phi)}}^2 > 0$, and therefore $\rho^q_{(\phi)}$ experiences neither tachyonic growth
nor parametric resonance. In contrast, $m_{\mathrm{eff,(h)}}^2 > 0$ before the end of inflation, but oscillates around the minimum after the end of inflation; as a result,
$\rho^q_{(h)}$ experiences parametric resonance. We now denote the completion of preheating as the time when $\rho_{\mathrm{inf}}=\rho_{\rm pre}=\rho^{(q)}_{(X)}$ (with $X$ being either $\phi$ or $h$), at a time $\mathcal{N}{\rm pre}$~\cite{Sfakianakis:2018lzf}.
It is clear from Fig.~\ref{plot:energy} that $\rho^q_{(\phi)}$ cannot provide successful preheating for any of the BPs, but preheating is possible
for BP$b$ via $\rho^q_{(h)}$. This is primarily due to the smaller $\xi_H$ in BP$a$. We find that for BP$b$, $\rho^q_{(h)} \approx \rho_{\rm inf}$ at $\mathcal{N} \approx 1.8$.
As discussed above, we assume that thermalization is complete immediately after preheating is completed~\cite{Cado:2024von}.
It is clear that preheating via the production of Higgs quanta is not possible for BP$a$; however, there is a subtlety, which we shall return to shortly.

We are now equipped with all quantities needed to evaluate the right side of Eq.~\eqref{eq:nstar}.
The  preheating temperature $T_{\rm pre}$ is evaluated equating $\rho_{\rm pre}$ to the  energy density of the thermal bath
\begin{align}
\rho_{\mathrm{inf}}\bigr|_{\mathcal{N}=\mathcal{N}_{\rm pre}}\equiv\rho_{\rm pre}=\frac{g_{\rm pre} \pi^2}{30} \; T_{\rm pre}^4,\label{eq:prehettemp}
\end{align}
with all relevant quantities are summarized in Table~\ref{table:reheating}.
\begin{table}[t!]
\begin{tabular}{|c |c| c| c| c | c | c| c | c |c| c| c| c}
    \hline
	BP  & preheating field(s)    & $\mathcal{N}_{\rm pre}$ & $\rho_{\rm pre} \; [M_{\rm P}^4]$ &  $T_{\rm pre}$ [GeV]\\
   \hline
        $a$ & --  &   -- &  --& -- \\
        $b$ & $h$  & 1.9   & $5.9 \times 10^{-14}$  &  $4.9 \times 10^{14}$ \\
	\hline
	\end{tabular}
	\caption{The details of preheating for both the benchmark points chosen for our analysis.}
	\label{table:reheating}
\end{table}
We find that $T_{\rm pre} \approx 4.9 \times 10^{14}$ GeV for BP$b$. This leads to an $e$-folding value of $55.39$ for the right-hand side of Eq.~\eqref{eq:nstar}. In contrast, as discussed above, the left-hand side yields $54.51$ $e$-foldings. This simply means that the matching is very close but not exact. Indeed, one can slightly readjust the model parameters $\xi_R$, $\xi_H$, and $\xi_c$, along with $\varphi(t_{\rm in})$ and $h_0(t_{\rm in})$, to achieve exact matching; however, we refrain from such adjustments here. This is primarily because we have not considered the decay of the inflaton and Higgs condensates, as well as the production of inflaton and Higgs quanta. Such decays will reduce $\rho_{\rm pre}$, leading to a lower $T_{\rm pre}$. We leave a detailed study for future work.

Although the preheating for BP$a$ via $\phi$ and $h$ particle production is incomplete, the associated longitudinal
gauge bosons (or, conversely, Goldstone bosons) may complete preheating even for BP$a$~\cite{Cado:2024von}. It has been
found that for pure $R^2$-Higgs inflation (i.e., without the $R^3$ term), with similar parameter values and $\xi_H \sim 1$,
the Goldstone bosons can preheat the Universe within $\mathcal{N} \simeq 3$ with a $T_{\rm pre} \approx 5 \times 10^{14}$ GeV~\cite{Cado:2024von}.
Inserting these numbers, we find that the left and right-hand sides of Eq.~\eqref{eq:nstar} give $53.66$ and $54.16$, respectively.
As discussed before, one can readjust the model parameters to achieve exact matching. A proper
estimation of preheating for both BP$a$ and BP$b$, including contributions from Goldstone and gauge
bosons, would require decomposing the Higgs field $\Phi$ in the Coulomb gauge instead of the unitary
gauge adopted here, as the latter becomes ill-defined during zero-crossings of the Higgs condensate
after the end of inflation~\cite{Sfakianakis:2018lzf,Cado:2024von}. This is deferred to a more detailed future publication.

\section{Summary and Outlook}
\label{sec:disc}
In this paper, we discuss how the addition of an $R^3$ term to $R^2$-Higgs inflation may account
for the observed high $n_s$ in the latest CMB+BAO data within the $\Lambda \rm{CDM}$ model~\cite{ACT:2025fju,SPT-3G:2025bzu}.
Considering two representative benchmark parameter sets, we find that $\xi_c \sim 10^{-13}$–$10^{-14}$ is
sufficient to explain the observed $n_s$, $\mathcal{A}_s$, and $r$ in the single-field-like regime.
We show that Higgs preheating plays a pivotal role in matching the reference CMB scale to
the inflationary scale with $\xi_H \gtrsim 10$. Smaller values, $\xi_H \sim 1$, can indeed provide
successful preheating, however, via the production of Goldstone bosons~\cite{Cado:2024von}, which is not considered here.
In addition, we have not considered the decay of the inflaton and Higgs condensate, nor the decay
of the produced quanta and their back-reaction on the background dynamics. This induces some uncertainties
in our result, but the model parameters can be readjusted to achieve exact matching of scales easily.
We also note that, in general, a negative value of $\xi_c$ can account for the observed value of $n_s$,
as discussed in Ref.~\cite{Addazi:2025qra}. In the case of $R^2$-Higgs inflation, the presence of the Higgs boson
introduces an additional field direction in the potential $V_E$. Consequently, a positive $\xi_c$ can also
reproduce the observed $n_s$ if one departs from the exact single-field regime, as shown above. In this first attempt, we have simply
assumed $\xi_c > 0$ and leave a more detailed analysis for future work.

We also identify further limitations here. It has been found that gauge boson production may
play a vital role in $R^2$-Higgs inflation for $\xi_H \gtrsim 1$\cite{Sfakianakis:2018lzf,Cado:2024von}.
However, to consider both Goldstone and gauge preheating, one must abandon the unitary gauge adopted here.
This is primarily because the unitary gauge becomes ill-defined at each zero crossing of the
Higgs condensate $h_0$\cite{Sfakianakis:2018lzf,Cado:2024von}.
Moreover, the produced gauge bosons may decay leptonically and inhibit the completion of gauge
preheating~\cite{Sfakianakis:2018lzf,Cado:2024von}. These factors introduce further uncertainties into our results.
To the best of our knowledge, such effects have not been considered in the literature for $R^2$-Higgs inflation following the ACT and SPT data.

At this point it is also useful to understand the role of BAO data within $\Lambda \rm{CDM}$ model. 
Note that the BAO data does not directly constraint the $n_s$ but, once they are combined with the CMB data correlated uncertainties shift the $n_s$ to 
a higher value. Indeed, the CMB data from ACT, SPT and Planck are consistent with each other and 
the combination of all CMB data i.e. the  CMB-SPA measurement found $n_s=0.9684 \pm 0.0030$ in $\Lambda \rm{CDM}$ model. This is consistent with Planck 2018
within $1\sigma$ and pure Starobinsky, Higgs and $R^2$-Higgs models are still best fit model to the data. It has been shown that 
the higher $n_s$  is due to the discrepancy between the CMB and BAO data and the discrepancy between BAO parameters and $n_s$ in
the CMB data within the $\Lambda \rm{CDM}$ model. While this apparent tension between CMB and BAO may well be artifact of unknown systematics,
it could well be an indication of new physics beyond $\Lambda \rm{CDM}$~\cite{Ferreira:2025lrd}.  Nonetheless, additional independent 
measurements of the CMB and BAO are required to resolve this tension. On the CMB side, the Simons Observatory 
is expected to achieve a precision of $\sigma(n_s) \sim 0.002$~\cite{SimonsObservatory:2025wwn}. 
BAO data from the Dark Energy Survey DES~\cite{DES:2024pwq}, and the
improved sensitivity of Euclid in measuring BAO~\cite{Euclid:2025dlg}, may also contribute significantly.
While we await more precise measurements,  within $\Lambda \rm{CDM}$ model, this high $n_s$ may indicate presence of $R^3$ term in $R^2$-Higgs inflation.
If confirmed in the future data, along with detection of $r$, this may offer exquisite information about quantum gravity.

\subsection*{Acknowledgments}
We thank Yann Cado and Evangelos I. Sfakianakis for useful discussion.

\renewcommand{\emph}{}
\bibliography{references}

\begin{thebibliography}{85}%
\makeatletter
\providecommand \@ifxundefined [1]{%
 \@ifx{#1\undefined}
}%
\providecommand \@ifnum [1]{%
 \ifnum #1\expandafter \@firstoftwo
 \else \expandafter \@secondoftwo
 \fi
}%
\providecommand \@ifx [1]{%
 \ifx #1\expandafter \@firstoftwo
 \else \expandafter \@secondoftwo
 \fi
}%
\providecommand \natexlab [1]{#1}%
\providecommand \enquote  [1]{``#1''}%
\providecommand \bibnamefont  [1]{#1}%
\providecommand \bibfnamefont [1]{#1}%
\providecommand \citenamefont [1]{#1}%
\providecommand \href@noop [0]{\@secondoftwo}%
\providecommand \href [0]{\begingroup \@sanitize@url \@href}%
\providecommand \@href[1]{\@@startlink{#1}\@@href}%
\providecommand \@@href[1]{\endgroup#1\@@endlink}%
\providecommand \@sanitize@url [0]{\catcode `\\12\catcode `\$12\catcode
  `\&12\catcode `\#12\catcode `\^12\catcode `\_12\catcode `\%12\relax}%
\providecommand \@@startlink[1]{}%
\providecommand \@@endlink[0]{}%
\providecommand \url  [0]{\begingroup\@sanitize@url \@url }%
\providecommand \@url [1]{\endgroup\@href {#1}{\urlprefix }}%
\providecommand \urlprefix  [0]{URL }%
\providecommand \Eprint [0]{\href }%
\providecommand \doibase [0]{https://doi.org/}%
\providecommand \selectlanguage [0]{\@gobble}%
\providecommand \bibinfo  [0]{\@secondoftwo}%
\providecommand \bibfield  [0]{\@secondoftwo}%
\providecommand \translation [1]{[#1]}%
\providecommand \BibitemOpen [0]{}%
\providecommand \bibitemStop [0]{}%
\providecommand \bibitemNoStop [0]{.\EOS\space}%
\providecommand \EOS [0]{\spacefactor3000\relax}%
\providecommand \BibitemShut  [1]{\csname bibitem#1\endcsname}%
\let\auto@bib@innerbib\@empty
\bibitem [{\citenamefont {Starobinsky}(1980)}]{Starobinsky:1980te}%
  \BibitemOpen
  \bibfield  {author} {\bibinfo {author} {\bibfnamefont {A.~A.}\ \bibnamefont
  {Starobinsky}},\ }\bibfield  {title} {\bibinfo {title} {{A New Type of
  Isotropic Cosmological Models Without Singularity}},\ }\href
  {https://doi.org/10.1016/0370-2693(80)90670-X} {\bibfield  {journal}
  {\bibinfo  {journal} {Phys. Lett. B}\ }\textbf {\bibinfo {volume} {91}},\
  \bibinfo {pages} {99} (\bibinfo {year} {1980})}\BibitemShut {NoStop}%
\bibitem [{\citenamefont {Sato}(1981)}]{Sato:1980yn}%
  \BibitemOpen
  \bibfield  {author} {\bibinfo {author} {\bibfnamefont {K.}~\bibnamefont
  {Sato}},\ }\bibfield  {title} {\bibinfo {title} {{First Order Phase
  Transition of a Vacuum and Expansion of the Universe}},\ }\href@noop {}
  {\bibfield  {journal} {\bibinfo  {journal} {Mon. Not. Roy. Astron. Soc.}\
  }\textbf {\bibinfo {volume} {195}},\ \bibinfo {pages} {467} (\bibinfo {year}
  {1981})}\BibitemShut {NoStop}%
\bibitem [{\citenamefont {Guth}(1981)}]{Guth:1980zm}%
  \BibitemOpen
  \bibfield  {author} {\bibinfo {author} {\bibfnamefont {A.~H.}\ \bibnamefont
  {Guth}},\ }\bibfield  {title} {\bibinfo {title} {{The Inflationary Universe:
  A Possible Solution to the Horizon and Flatness Problems}},\ }\href
  {https://doi.org/10.1103/PhysRevD.23.347} {\bibfield  {journal} {\bibinfo
  {journal} {Phys. Rev. D}\ }\textbf {\bibinfo {volume} {23}},\ \bibinfo
  {pages} {347} (\bibinfo {year} {1981})}\BibitemShut {NoStop}%
\bibitem [{\citenamefont {Hinshaw}\ \emph {et~al.}(2013)\citenamefont {Hinshaw}
  \emph {et~al.}}]{WMAP:2012nax}%
  \BibitemOpen
  \bibfield  {author} {\bibinfo {author} {\bibfnamefont {G.}~\bibnamefont
  {Hinshaw}} \emph {et~al.} (\bibinfo {collaboration} {WMAP}),\ }\bibfield
  {title} {\bibinfo {title} {{Nine-Year Wilkinson Microwave Anisotropy Probe
  (WMAP) Observations: Cosmological Parameter Results}},\ }\href
  {https://doi.org/10.1088/0067-0049/208/2/19} {\bibfield  {journal} {\bibinfo
  {journal} {Astrophys. J. Suppl.}\ }\textbf {\bibinfo {volume} {208}},\
  \bibinfo {pages} {19} (\bibinfo {year} {2013})},\ \Eprint
  {https://arxiv.org/abs/1212.5226} {arXiv:1212.5226 [astro-ph.CO]}
  \BibitemShut {NoStop}%
\bibitem [{\citenamefont {Akrami}\ \emph {et~al.}(2020)\citenamefont {Akrami}
  \emph {et~al.}}]{Planck:2018jri}%
  \BibitemOpen
  \bibfield  {author} {\bibinfo {author} {\bibfnamefont {Y.}~\bibnamefont
  {Akrami}} \emph {et~al.} (\bibinfo {collaboration} {Planck}),\ }\bibfield
  {title} {\bibinfo {title} {{Planck 2018 results. X. Constraints on
  inflation}},\ }\href {https://doi.org/10.1051/0004-6361/201833887} {\bibfield
   {journal} {\bibinfo  {journal} {Astron. Astrophys.}\ }\textbf {\bibinfo
  {volume} {641}},\ \bibinfo {pages} {A10} (\bibinfo {year} {2020})},\ \Eprint
  {https://arxiv.org/abs/1807.06211} {arXiv:1807.06211 [astro-ph.CO]}
  \BibitemShut {NoStop}%
\bibitem [{\citenamefont {Ade}\ \emph {et~al.}(2021)\citenamefont {Ade} \emph
  {et~al.}}]{BICEP:2021xfz}%
  \BibitemOpen
  \bibfield  {author} {\bibinfo {author} {\bibfnamefont {P.~A.~R.}\
  \bibnamefont {Ade}} \emph {et~al.} (\bibinfo {collaboration} {BICEP, Keck}),\
  }\bibfield  {title} {\bibinfo {title} {{Improved Constraints on Primordial
  Gravitational Waves using Planck, WMAP, and BICEP/Keck Observations through
  the 2018 Observing Season}},\ }\href
  {https://doi.org/10.1103/PhysRevLett.127.151301} {\bibfield  {journal}
  {\bibinfo  {journal} {Phys. Rev. Lett.}\ }\textbf {\bibinfo {volume} {127}},\
  \bibinfo {pages} {151301} (\bibinfo {year} {2021})},\ \Eprint
  {https://arxiv.org/abs/2110.00483} {arXiv:2110.00483 [astro-ph.CO]}
  \BibitemShut {NoStop}%
\bibitem [{\citenamefont {Ade}\ \emph {et~al.}(2019)\citenamefont {Ade} \emph
  {et~al.}}]{SimonsObservatory:2018koc}%
  \BibitemOpen
  \bibfield  {author} {\bibinfo {author} {\bibfnamefont {P.}~\bibnamefont
  {Ade}} \emph {et~al.} (\bibinfo {collaboration} {Simons Observatory}),\
  }\bibfield  {title} {\bibinfo {title} {{The Simons Observatory: Science goals
  and forecasts}},\ }\href {https://doi.org/10.1088/1475-7516/2019/02/056}
  {\bibfield  {journal} {\bibinfo  {journal} {JCAP}\ }\textbf {\bibinfo
  {volume} {02}},\ \bibinfo {pages} {056}},\ \Eprint
  {https://arxiv.org/abs/1808.07445} {arXiv:1808.07445 [astro-ph.CO]}
  \BibitemShut {NoStop}%
\bibitem [{\citenamefont {Allys}\ \emph {et~al.}(2023)\citenamefont {Allys}
  \emph {et~al.}}]{LiteBIRD:2022cnt}%
  \BibitemOpen
  \bibfield  {author} {\bibinfo {author} {\bibfnamefont {E.}~\bibnamefont
  {Allys}} \emph {et~al.} (\bibinfo {collaboration} {LiteBIRD}),\ }\bibfield
  {title} {\bibinfo {title} {{Probing Cosmic Inflation with the LiteBIRD Cosmic
  Microwave Background Polarization Survey}},\ }\href
  {https://doi.org/10.1093/ptep/ptac150} {\bibfield  {journal} {\bibinfo
  {journal} {PTEP}\ }\textbf {\bibinfo {volume} {2023}},\ \bibinfo {pages}
  {042F01} (\bibinfo {year} {2023})},\ \Eprint
  {https://arxiv.org/abs/2202.02773} {arXiv:2202.02773 [astro-ph.IM]}
  \BibitemShut {NoStop}%
\bibitem [{\citenamefont {Starobinsky}(1983)}]{Starobinsky:1983zz}%
  \BibitemOpen
  \bibfield  {author} {\bibinfo {author} {\bibfnamefont {A.~A.}\ \bibnamefont
  {Starobinsky}},\ }\bibfield  {title} {\bibinfo {title} {{The Perturbation
  Spectrum Evolving from a Nonsingular Initially De-Sitter Cosmology and the
  Microwave Background Anisotropy}},\ }\href@noop {} {\bibfield  {journal}
  {\bibinfo  {journal} {Sov. Astron. Lett.}\ }\textbf {\bibinfo {volume} {9}},\
  \bibinfo {pages} {302} (\bibinfo {year} {1983})}\BibitemShut {NoStop}%
\bibitem [{\citenamefont {Vilenkin}(1985)}]{Vilenkin:1985md}%
  \BibitemOpen
  \bibfield  {author} {\bibinfo {author} {\bibfnamefont {A.}~\bibnamefont
  {Vilenkin}},\ }\bibfield  {title} {\bibinfo {title} {{Classical and Quantum
  Cosmology of the Starobinsky Inflationary Model}},\ }\href
  {https://doi.org/10.1103/PhysRevD.32.2511} {\bibfield  {journal} {\bibinfo
  {journal} {Phys. Rev. D}\ }\textbf {\bibinfo {volume} {32}},\ \bibinfo
  {pages} {2511} (\bibinfo {year} {1985})}\BibitemShut {NoStop}%
\bibitem [{\citenamefont {Mijic}\ \emph {et~al.}(1986)\citenamefont {Mijic},
  \citenamefont {Morris},\ and\ \citenamefont {Suen}}]{Mijic:1986iv}%
  \BibitemOpen
  \bibfield  {author} {\bibinfo {author} {\bibfnamefont {M.~B.}\ \bibnamefont
  {Mijic}}, \bibinfo {author} {\bibfnamefont {M.~S.}\ \bibnamefont {Morris}},\
  and\ \bibinfo {author} {\bibfnamefont {W.-M.}\ \bibnamefont {Suen}},\
  }\bibfield  {title} {\bibinfo {title} {{The $R^2$ Cosmology: Inflation
  Without a Phase Transition}},\ }\href
  {https://doi.org/10.1103/PhysRevD.34.2934} {\bibfield  {journal} {\bibinfo
  {journal} {Phys. Rev. D}\ }\textbf {\bibinfo {volume} {34}},\ \bibinfo
  {pages} {2934} (\bibinfo {year} {1986})}\BibitemShut {NoStop}%
\bibitem [{\citenamefont {Maeda}(1988)}]{Maeda:1987xf}%
  \BibitemOpen
  \bibfield  {author} {\bibinfo {author} {\bibfnamefont {K.-i.}\ \bibnamefont
  {Maeda}},\ }\bibfield  {title} {\bibinfo {title} {{Inflation as a Transient
  Attractor in $R^2$ Cosmology}},\ }\href
  {https://doi.org/10.1103/PhysRevD.37.858} {\bibfield  {journal} {\bibinfo
  {journal} {Phys. Rev. D}\ }\textbf {\bibinfo {volume} {37}},\ \bibinfo
  {pages} {858} (\bibinfo {year} {1988})}\BibitemShut {NoStop}%
\bibitem [{\citenamefont {Louis}\ \emph {et~al.}(2025)\citenamefont {Louis}
  \emph {et~al.}}]{ACT:2025fju}%
  \BibitemOpen
  \bibfield  {author} {\bibinfo {author} {\bibfnamefont {T.}~\bibnamefont
  {Louis}} \emph {et~al.} (\bibinfo {collaboration} {ACT}),\ }\bibfield
  {title} {\bibinfo {title} {{The Atacama Cosmology Telescope: DR6 Power
  Spectra, Likelihoods and $Λ$CDM Parameters}},\ }\href@noop {} {\  (\bibinfo
  {year} {2025})},\ \Eprint {https://arxiv.org/abs/2503.14452}
  {arXiv:2503.14452 [astro-ph.CO]} \BibitemShut {NoStop}%
\bibitem [{\citenamefont {Camphuis}\ \emph {et~al.}(2025)\citenamefont
  {Camphuis} \emph {et~al.}}]{SPT-3G:2025bzu}%
  \BibitemOpen
  \bibfield  {author} {\bibinfo {author} {\bibfnamefont {E.}~\bibnamefont
  {Camphuis}} \emph {et~al.} (\bibinfo {collaboration} {SPT-3G}),\ }\bibfield
  {title} {\bibinfo {title} {{SPT-3G D1: CMB temperature and polarization power
  spectra and cosmology from 2019 and 2020 observations of the SPT-3G Main
  field}},\ }\href@noop {} {\  (\bibinfo {year} {2025})},\ \Eprint
  {https://arxiv.org/abs/2506.20707} {arXiv:2506.20707 [astro-ph.CO]}
  \BibitemShut {NoStop}%
\bibitem [{\citenamefont {Abdul~Karim}\ \emph {et~al.}(2025)\citenamefont
  {Abdul~Karim} \emph {et~al.}}]{DESI:2025zgx}%
  \BibitemOpen
  \bibfield  {author} {\bibinfo {author} {\bibfnamefont {M.}~\bibnamefont
  {Abdul~Karim}} \emph {et~al.} (\bibinfo {collaboration} {DESI}),\ }\bibfield
  {title} {\bibinfo {title} {{DESI DR2 Results II: Measurements of Baryon
  Acoustic Oscillations and Cosmological Constraints}},\ }\href@noop {} {\
  (\bibinfo {year} {2025})},\ \Eprint {https://arxiv.org/abs/2503.14738}
  {arXiv:2503.14738 [astro-ph.CO]} \BibitemShut {NoStop}%
\bibitem [{\citenamefont {Salvio}\ and\ \citenamefont
  {Mazumdar}(2015)}]{Salvio:2015kka}%
  \BibitemOpen
  \bibfield  {author} {\bibinfo {author} {\bibfnamefont {A.}~\bibnamefont
  {Salvio}}\ and\ \bibinfo {author} {\bibfnamefont {A.}~\bibnamefont
  {Mazumdar}},\ }\bibfield  {title} {\bibinfo {title} {{Classical and Quantum
  Initial Conditions for Higgs Inflation}},\ }\href
  {https://doi.org/10.1016/j.physletb.2015.09.020} {\bibfield  {journal}
  {\bibinfo  {journal} {Phys. Lett. B}\ }\textbf {\bibinfo {volume} {750}},\
  \bibinfo {pages} {194} (\bibinfo {year} {2015})},\ \Eprint
  {https://arxiv.org/abs/1506.07520} {arXiv:1506.07520 [hep-ph]} \BibitemShut
  {NoStop}%
\bibitem [{\citenamefont {Ema}(2017)}]{Ema:2017rqn}%
  \BibitemOpen
  \bibfield  {author} {\bibinfo {author} {\bibfnamefont {Y.}~\bibnamefont
  {Ema}},\ }\bibfield  {title} {\bibinfo {title} {{Higgs Scalaron Mixed
  Inflation}},\ }\href {https://doi.org/10.1016/j.physletb.2017.04.060}
  {\bibfield  {journal} {\bibinfo  {journal} {Phys. Lett. B}\ }\textbf
  {\bibinfo {volume} {770}},\ \bibinfo {pages} {403} (\bibinfo {year}
  {2017})},\ \Eprint {https://arxiv.org/abs/1701.07665} {arXiv:1701.07665
  [hep-ph]} \BibitemShut {NoStop}%
\bibitem [{\citenamefont {Pi}\ \emph {et~al.}(2018)\citenamefont {Pi},
  \citenamefont {Zhang}, \citenamefont {Huang},\ and\ \citenamefont
  {Sasaki}}]{Pi:2017gih}%
  \BibitemOpen
  \bibfield  {author} {\bibinfo {author} {\bibfnamefont {S.}~\bibnamefont
  {Pi}}, \bibinfo {author} {\bibfnamefont {Y.-l.}\ \bibnamefont {Zhang}},
  \bibinfo {author} {\bibfnamefont {Q.-G.}\ \bibnamefont {Huang}},\ and\
  \bibinfo {author} {\bibfnamefont {M.}~\bibnamefont {Sasaki}},\ }\bibfield
  {title} {\bibinfo {title} {{Scalaron from $R^2$-gravity as a heavy field}},\
  }\href {https://doi.org/10.1088/1475-7516/2018/05/042} {\bibfield  {journal}
  {\bibinfo  {journal} {JCAP}\ }\textbf {\bibinfo {volume} {05}},\ \bibinfo
  {pages} {042}},\ \Eprint {https://arxiv.org/abs/1712.09896} {arXiv:1712.09896
  [astro-ph.CO]} \BibitemShut {NoStop}%
\bibitem [{\citenamefont {Gorbunov}\ and\ \citenamefont
  {Tokareva}(2019)}]{Gorbunov:2018llf}%
  \BibitemOpen
  \bibfield  {author} {\bibinfo {author} {\bibfnamefont {D.}~\bibnamefont
  {Gorbunov}}\ and\ \bibinfo {author} {\bibfnamefont {A.}~\bibnamefont
  {Tokareva}},\ }\bibfield  {title} {\bibinfo {title} {{Scalaron the healer:
  removing the strong-coupling in the Higgs- and Higgs-dilaton inflations}},\
  }\href {https://doi.org/10.1016/j.physletb.2018.11.015} {\bibfield  {journal}
  {\bibinfo  {journal} {Phys. Lett. B}\ }\textbf {\bibinfo {volume} {788}},\
  \bibinfo {pages} {37} (\bibinfo {year} {2019})},\ \Eprint
  {https://arxiv.org/abs/1807.02392} {arXiv:1807.02392 [hep-ph]} \BibitemShut
  {NoStop}%
\bibitem [{\citenamefont {Gundhi}\ and\ \citenamefont
  {Steinwachs}(2020)}]{Gundhi:2018wyz}%
  \BibitemOpen
  \bibfield  {author} {\bibinfo {author} {\bibfnamefont {A.}~\bibnamefont
  {Gundhi}}\ and\ \bibinfo {author} {\bibfnamefont {C.~F.}\ \bibnamefont
  {Steinwachs}},\ }\bibfield  {title} {\bibinfo {title} {{Scalaron-Higgs
  inflation}},\ }\href {https://doi.org/10.1016/j.nuclphysb.2020.114989}
  {\bibfield  {journal} {\bibinfo  {journal} {Nucl. Phys. B}\ }\textbf
  {\bibinfo {volume} {954}},\ \bibinfo {pages} {114989} (\bibinfo {year}
  {2020})},\ \Eprint {https://arxiv.org/abs/1810.10546} {arXiv:1810.10546
  [hep-th]} \BibitemShut {NoStop}%
\bibitem [{\citenamefont {He}\ \emph {et~al.}(2019)\citenamefont {He},
  \citenamefont {Jinno}, \citenamefont {Kamada}, \citenamefont {Park},
  \citenamefont {Starobinsky},\ and\ \citenamefont {Yokoyama}}]{He:2018mgb}%
  \BibitemOpen
  \bibfield  {author} {\bibinfo {author} {\bibfnamefont {M.}~\bibnamefont
  {He}}, \bibinfo {author} {\bibfnamefont {R.}~\bibnamefont {Jinno}}, \bibinfo
  {author} {\bibfnamefont {K.}~\bibnamefont {Kamada}}, \bibinfo {author}
  {\bibfnamefont {S.~C.}\ \bibnamefont {Park}}, \bibinfo {author}
  {\bibfnamefont {A.~A.}\ \bibnamefont {Starobinsky}},\ and\ \bibinfo {author}
  {\bibfnamefont {J.}~\bibnamefont {Yokoyama}},\ }\bibfield  {title} {\bibinfo
  {title} {{On the violent preheating in the mixed Higgs-$R^2$ inflationary
  model}},\ }\href {https://doi.org/10.1016/j.physletb.2019.02.008} {\bibfield
  {journal} {\bibinfo  {journal} {Phys. Lett. B}\ }\textbf {\bibinfo {volume}
  {791}},\ \bibinfo {pages} {36} (\bibinfo {year} {2019})},\ \Eprint
  {https://arxiv.org/abs/1812.10099} {arXiv:1812.10099 [hep-ph]} \BibitemShut
  {NoStop}%
\bibitem [{\citenamefont {He}\ \emph {et~al.}(2018)\citenamefont {He},
  \citenamefont {Starobinsky},\ and\ \citenamefont {Yokoyama}}]{He:2018gyf}%
  \BibitemOpen
  \bibfield  {author} {\bibinfo {author} {\bibfnamefont {M.}~\bibnamefont
  {He}}, \bibinfo {author} {\bibfnamefont {A.~A.}\ \bibnamefont
  {Starobinsky}},\ and\ \bibinfo {author} {\bibfnamefont {J.}~\bibnamefont
  {Yokoyama}},\ }\bibfield  {title} {\bibinfo {title} {{Inflation in the mixed
  Higgs-$R^2$ model}},\ }\href {https://doi.org/10.1088/1475-7516/2018/05/064}
  {\bibfield  {journal} {\bibinfo  {journal} {JCAP}\ }\textbf {\bibinfo
  {volume} {05}},\ \bibinfo {pages} {064}},\ \Eprint
  {https://arxiv.org/abs/1804.00409} {arXiv:1804.00409 [astro-ph.CO]}
  \BibitemShut {NoStop}%
\bibitem [{\citenamefont {Cheong}\ \emph {et~al.}(2021)\citenamefont {Cheong},
  \citenamefont {Lee},\ and\ \citenamefont {Park}}]{Cheong:2019vzl}%
  \BibitemOpen
  \bibfield  {author} {\bibinfo {author} {\bibfnamefont {D.~Y.}\ \bibnamefont
  {Cheong}}, \bibinfo {author} {\bibfnamefont {S.~M.}\ \bibnamefont {Lee}},\
  and\ \bibinfo {author} {\bibfnamefont {S.~C.}\ \bibnamefont {Park}},\
  }\bibfield  {title} {\bibinfo {title} {{Primordial black holes in Higgs-$R^2$
  inflation as the whole of dark matter}},\ }\href
  {https://doi.org/10.1088/1475-7516/2021/01/032} {\bibfield  {journal}
  {\bibinfo  {journal} {JCAP}\ }\textbf {\bibinfo {volume} {01}},\ \bibinfo
  {pages} {032}},\ \Eprint {https://arxiv.org/abs/1912.12032} {arXiv:1912.12032
  [hep-ph]} \BibitemShut {NoStop}%
\bibitem [{\citenamefont {Canko}\ \emph {et~al.}(2020)\citenamefont {Canko},
  \citenamefont {Gialamas},\ and\ \citenamefont {Kodaxis}}]{Canko:2019mud}%
  \BibitemOpen
  \bibfield  {author} {\bibinfo {author} {\bibfnamefont {D.~D.}\ \bibnamefont
  {Canko}}, \bibinfo {author} {\bibfnamefont {I.~D.}\ \bibnamefont
  {Gialamas}},\ and\ \bibinfo {author} {\bibfnamefont {G.~P.}\ \bibnamefont
  {Kodaxis}},\ }\bibfield  {title} {\bibinfo {title} {{A simple
  $F(\mathcal{R},\phi )$ deformation of Starobinsky inflationary model}},\
  }\href {https://doi.org/10.1140/epjc/s10052-020-8025-4} {\bibfield  {journal}
  {\bibinfo  {journal} {Eur. Phys. J. C}\ }\textbf {\bibinfo {volume} {80}},\
  \bibinfo {pages} {458} (\bibinfo {year} {2020})},\ \Eprint
  {https://arxiv.org/abs/1901.06296} {arXiv:1901.06296 [hep-th]} \BibitemShut
  {NoStop}%
\bibitem [{\citenamefont {He}\ \emph {et~al.}(2021)\citenamefont {He},
  \citenamefont {Jinno}, \citenamefont {Kamada}, \citenamefont {Starobinsky},\
  and\ \citenamefont {Yokoyama}}]{He:2020ivk}%
  \BibitemOpen
  \bibfield  {author} {\bibinfo {author} {\bibfnamefont {M.}~\bibnamefont
  {He}}, \bibinfo {author} {\bibfnamefont {R.}~\bibnamefont {Jinno}}, \bibinfo
  {author} {\bibfnamefont {K.}~\bibnamefont {Kamada}}, \bibinfo {author}
  {\bibfnamefont {A.~A.}\ \bibnamefont {Starobinsky}},\ and\ \bibinfo {author}
  {\bibfnamefont {J.}~\bibnamefont {Yokoyama}},\ }\bibfield  {title} {\bibinfo
  {title} {{Occurrence of tachyonic preheating in the mixed Higgs-R$^2$
  model}},\ }\href {https://doi.org/10.1088/1475-7516/2021/01/066} {\bibfield
  {journal} {\bibinfo  {journal} {JCAP}\ }\textbf {\bibinfo {volume} {01}},\
  \bibinfo {pages} {066}},\ \Eprint {https://arxiv.org/abs/2007.10369}
  {arXiv:2007.10369 [hep-ph]} \BibitemShut {NoStop}%
\bibitem [{\citenamefont {He}(2021)}]{He:2020qcb}%
  \BibitemOpen
  \bibfield  {author} {\bibinfo {author} {\bibfnamefont {M.}~\bibnamefont
  {He}},\ }\bibfield  {title} {\bibinfo {title} {{Perturbative Reheating in the
  Mixed Higgs-$R^2$ Model}},\ }\href
  {https://doi.org/10.1088/1475-7516/2021/05/021} {\bibfield  {journal}
  {\bibinfo  {journal} {JCAP}\ }\textbf {\bibinfo {volume} {05}},\ \bibinfo
  {pages} {021}},\ \Eprint {https://arxiv.org/abs/2010.11717} {arXiv:2010.11717
  [hep-ph]} \BibitemShut {NoStop}%
\bibitem [{\citenamefont {Pineda}\ and\ \citenamefont
  {Pimentel}(2024)}]{Pineda:2024prs}%
  \BibitemOpen
  \bibfield  {author} {\bibinfo {author} {\bibfnamefont {F.}~\bibnamefont
  {Pineda}}\ and\ \bibinfo {author} {\bibfnamefont {L.~O.}\ \bibnamefont
  {Pimentel}},\ }\bibfield  {title} {\bibinfo {title} {{Non-perturbative
  approach for scalar particle production in Higgs-$R^2$ inflation}},\
  }\href@noop {} {\  (\bibinfo {year} {2024})},\ \Eprint
  {https://arxiv.org/abs/2409.09999} {arXiv:2409.09999 [hep-th]} \BibitemShut
  {NoStop}%
\bibitem [{\citenamefont {Copeland}\ \emph {et~al.}(2015)\citenamefont
  {Copeland}, \citenamefont {Rahmede},\ and\ \citenamefont
  {Saltas}}]{Copeland:2013vva}%
  \BibitemOpen
  \bibfield  {author} {\bibinfo {author} {\bibfnamefont {E.~J.}\ \bibnamefont
  {Copeland}}, \bibinfo {author} {\bibfnamefont {C.}~\bibnamefont {Rahmede}},\
  and\ \bibinfo {author} {\bibfnamefont {I.~D.}\ \bibnamefont {Saltas}},\
  }\bibfield  {title} {\bibinfo {title} {{Asymptotically Safe Starobinsky
  Inflation}},\ }\href {https://doi.org/10.1103/PhysRevD.91.103530} {\bibfield
  {journal} {\bibinfo  {journal} {Phys. Rev. D}\ }\textbf {\bibinfo {volume}
  {91}},\ \bibinfo {pages} {103530} (\bibinfo {year} {2015})},\ \Eprint
  {https://arxiv.org/abs/1311.0881} {arXiv:1311.0881 [gr-qc]} \BibitemShut
  {NoStop}%
\bibitem [{\citenamefont {Saltas}(2016)}]{Saltas:2015vsc}%
  \BibitemOpen
  \bibfield  {author} {\bibinfo {author} {\bibfnamefont {I.~D.}\ \bibnamefont
  {Saltas}},\ }\bibfield  {title} {\bibinfo {title} {{Higgs inflation and
  quantum gravity: An exact renormalisation group approach}},\ }\href
  {https://doi.org/10.1088/1475-7516/2016/02/048} {\bibfield  {journal}
  {\bibinfo  {journal} {JCAP}\ }\textbf {\bibinfo {volume} {02}},\ \bibinfo
  {pages} {048}},\ \Eprint {https://arxiv.org/abs/1512.06134} {arXiv:1512.06134
  [hep-th]} \BibitemShut {NoStop}%
\bibitem [{\citenamefont {de~la Cruz-Dombriz}\ \emph
  {et~al.}(2016)\citenamefont {de~la Cruz-Dombriz}, \citenamefont {Elizalde},
  \citenamefont {Odintsov},\ and\ \citenamefont
  {S{\'a}ez-G{\'o}mez}}]{delaCruz-Dombriz:2016bjj}%
  \BibitemOpen
  \bibfield  {author} {\bibinfo {author} {\bibfnamefont {A.}~\bibnamefont
  {de~la Cruz-Dombriz}}, \bibinfo {author} {\bibfnamefont {E.}~\bibnamefont
  {Elizalde}}, \bibinfo {author} {\bibfnamefont {S.~D.}\ \bibnamefont
  {Odintsov}},\ and\ \bibinfo {author} {\bibfnamefont {D.}~\bibnamefont
  {S{\'a}ez-G{\'o}mez}},\ }\bibfield  {title} {\bibinfo {title} {{Spotting
  deviations from R$^2$ inflation}},\ }\href
  {https://doi.org/10.1088/1475-7516/2016/05/060} {\bibfield  {journal}
  {\bibinfo  {journal} {JCAP}\ }\textbf {\bibinfo {volume} {05}},\ \bibinfo
  {pages} {060}},\ \Eprint {https://arxiv.org/abs/1603.05537} {arXiv:1603.05537
  [gr-qc]} \BibitemShut {NoStop}%
\bibitem [{\citenamefont {Odintsov}\ \emph {et~al.}(2021)\citenamefont
  {Odintsov}, \citenamefont {Oikonomou},\ and\ \citenamefont
  {Fronimos}}]{Odintsov:2020ilr}%
  \BibitemOpen
  \bibfield  {author} {\bibinfo {author} {\bibfnamefont {S.~D.}\ \bibnamefont
  {Odintsov}}, \bibinfo {author} {\bibfnamefont {V.~K.}\ \bibnamefont
  {Oikonomou}},\ and\ \bibinfo {author} {\bibfnamefont {F.~P.}\ \bibnamefont
  {Fronimos}},\ }\bibfield  {title} {\bibinfo {title} {{Canonical scalar field
  inflation with string and $R^2$ -corrections}},\ }\href
  {https://doi.org/10.1016/j.aop.2020.168359} {\bibfield  {journal} {\bibinfo
  {journal} {Annals Phys.}\ }\textbf {\bibinfo {volume} {424}},\ \bibinfo
  {pages} {168359} (\bibinfo {year} {2021})},\ \Eprint
  {https://arxiv.org/abs/2011.08680} {arXiv:2011.08680 [gr-qc]} \BibitemShut
  {NoStop}%
\bibitem [{\citenamefont {Gialamas}\ \emph
  {et~al.}(2025{\natexlab{a}})\citenamefont {Gialamas}, \citenamefont {Karam},
  \citenamefont {Racioppi},\ and\ \citenamefont {Raidal}}]{Gialamas:2025kef}%
  \BibitemOpen
  \bibfield  {author} {\bibinfo {author} {\bibfnamefont {I.~D.}\ \bibnamefont
  {Gialamas}}, \bibinfo {author} {\bibfnamefont {A.}~\bibnamefont {Karam}},
  \bibinfo {author} {\bibfnamefont {A.}~\bibnamefont {Racioppi}},\ and\
  \bibinfo {author} {\bibfnamefont {M.}~\bibnamefont {Raidal}},\ }\bibfield
  {title} {\bibinfo {title} {{Has ACT measured radiative corrections to the
  tree-level Higgs-like inflation?}},\ }\href@noop {} {\  (\bibinfo {year}
  {2025}{\natexlab{a}})},\ \Eprint {https://arxiv.org/abs/2504.06002}
  {arXiv:2504.06002 [astro-ph.CO]} \BibitemShut {NoStop}%
\bibitem [{\citenamefont {Drees}\ and\ \citenamefont
  {Xu}(2025)}]{Drees:2025ngb}%
  \BibitemOpen
  \bibfield  {author} {\bibinfo {author} {\bibfnamefont {M.}~\bibnamefont
  {Drees}}\ and\ \bibinfo {author} {\bibfnamefont {Y.}~\bibnamefont {Xu}},\
  }\bibfield  {title} {\bibinfo {title} {{Refined predictions for Starobinsky
  inflation and post-inflationary constraints in light of ACT}},\ }\href
  {https://doi.org/10.1016/j.physletb.2025.139612} {\bibfield  {journal}
  {\bibinfo  {journal} {Phys. Lett. B}\ }\textbf {\bibinfo {volume} {867}},\
  \bibinfo {pages} {139612} (\bibinfo {year} {2025})},\ \Eprint
  {https://arxiv.org/abs/2504.20757} {arXiv:2504.20757 [astro-ph.CO]}
  \BibitemShut {NoStop}%
\bibitem [{\citenamefont {Zharov}\ \emph
  {et~al.}(2025{\natexlab{a}})\citenamefont {Zharov}, \citenamefont {Sobol},\
  and\ \citenamefont {Vilchinskii}}]{Zharov:2025evb}%
  \BibitemOpen
  \bibfield  {author} {\bibinfo {author} {\bibfnamefont {D.~S.}\ \bibnamefont
  {Zharov}}, \bibinfo {author} {\bibfnamefont {O.~O.}\ \bibnamefont {Sobol}},\
  and\ \bibinfo {author} {\bibfnamefont {S.~I.}\ \bibnamefont {Vilchinskii}},\
  }\bibfield  {title} {\bibinfo {title} {{Reheating ACTs on Starobinsky and
  Higgs inflation}},\ }\href@noop {} {\  (\bibinfo {year}
  {2025}{\natexlab{a}})},\ \Eprint {https://arxiv.org/abs/2505.01129}
  {arXiv:2505.01129 [astro-ph.CO]} \BibitemShut {NoStop}%
\bibitem [{\citenamefont {Liu}\ \emph {et~al.}(2025)\citenamefont {Liu},
  \citenamefont {Yi},\ and\ \citenamefont {Gong}}]{Liu:2025qca}%
  \BibitemOpen
  \bibfield  {author} {\bibinfo {author} {\bibfnamefont {L.}~\bibnamefont
  {Liu}}, \bibinfo {author} {\bibfnamefont {Z.}~\bibnamefont {Yi}},\ and\
  \bibinfo {author} {\bibfnamefont {Y.}~\bibnamefont {Gong}},\ }\bibfield
  {title} {\bibinfo {title} {{Reconciling Higgs Inflation with ACT Observations
  through Reheating}},\ }\href@noop {} {\  (\bibinfo {year} {2025})},\ \Eprint
  {https://arxiv.org/abs/2505.02407} {arXiv:2505.02407 [astro-ph.CO]}
  \BibitemShut {NoStop}%
\bibitem [{\citenamefont {Haque}\ \emph {et~al.}(2025)\citenamefont {Haque},
  \citenamefont {Pal},\ and\ \citenamefont {Paul}}]{Haque:2025uis}%
  \BibitemOpen
  \bibfield  {author} {\bibinfo {author} {\bibfnamefont {M.~R.}\ \bibnamefont
  {Haque}}, \bibinfo {author} {\bibfnamefont {S.}~\bibnamefont {Pal}},\ and\
  \bibinfo {author} {\bibfnamefont {D.}~\bibnamefont {Paul}},\ }\bibfield
  {title} {\bibinfo {title} {{Improved Predictions on Higgs-Starobinsky
  Inflation and Reheating with ACT DR6 and Primordial Gravitational Waves}},\
  }\href@noop {} {\  (\bibinfo {year} {2025})},\ \Eprint
  {https://arxiv.org/abs/2505.04615} {arXiv:2505.04615 [astro-ph.CO]}
  \BibitemShut {NoStop}%
\bibitem [{\citenamefont {Yogesh}\ \emph {et~al.}(2025)\citenamefont {Yogesh},
  \citenamefont {Mohammadi}, \citenamefont {Wu},\ and\ \citenamefont
  {Zhu}}]{Yogesh:2025wak}%
  \BibitemOpen
  \bibfield  {author} {\bibinfo {author} {\bibnamefont {Yogesh}}, \bibinfo
  {author} {\bibfnamefont {A.}~\bibnamefont {Mohammadi}}, \bibinfo {author}
  {\bibfnamefont {Q.}~\bibnamefont {Wu}},\ and\ \bibinfo {author}
  {\bibfnamefont {T.}~\bibnamefont {Zhu}},\ }\bibfield  {title} {\bibinfo
  {title} {{Starobinsky like inflation and EGB Gravity in the light of ACT}},\
  }\href@noop {} {\  (\bibinfo {year} {2025})},\ \Eprint
  {https://arxiv.org/abs/2505.05363} {arXiv:2505.05363 [astro-ph.CO]}
  \BibitemShut {NoStop}%
\bibitem [{\citenamefont {Addazi}\ \emph {et~al.}(2025)\citenamefont {Addazi},
  \citenamefont {Aldabergenov},\ and\ \citenamefont {Ketov}}]{Addazi:2025qra}%
  \BibitemOpen
  \bibfield  {author} {\bibinfo {author} {\bibfnamefont {A.}~\bibnamefont
  {Addazi}}, \bibinfo {author} {\bibfnamefont {Y.}~\bibnamefont
  {Aldabergenov}},\ and\ \bibinfo {author} {\bibfnamefont {S.~V.}\ \bibnamefont
  {Ketov}},\ }\bibfield  {title} {\bibinfo {title} {{Curvature corrections to
  Starobinsky inflation can explain the ACT results}},\ }\href@noop {} {\
  (\bibinfo {year} {2025})},\ \Eprint {https://arxiv.org/abs/2505.10305}
  {arXiv:2505.10305 [gr-qc]} \BibitemShut {NoStop}%
\bibitem [{\citenamefont {Kallosh}\ and\ \citenamefont
  {Linde}(2025)}]{Kallosh:2025ijd}%
  \BibitemOpen
  \bibfield  {author} {\bibinfo {author} {\bibfnamefont {R.}~\bibnamefont
  {Kallosh}}\ and\ \bibinfo {author} {\bibfnamefont {A.}~\bibnamefont
  {Linde}},\ }\bibfield  {title} {\bibinfo {title} {{On the Present Status of
  Inflationary Cosmology}},\ }\href@noop {} {\  (\bibinfo {year} {2025})},\
  \Eprint {https://arxiv.org/abs/2505.13646} {arXiv:2505.13646 [hep-th]}
  \BibitemShut {NoStop}%
\bibitem [{\citenamefont {Gialamas}\ \emph
  {et~al.}(2025{\natexlab{b}})\citenamefont {Gialamas}, \citenamefont
  {Katsoulas},\ and\ \citenamefont {Tamvakis}}]{Gialamas:2025ofz}%
  \BibitemOpen
  \bibfield  {author} {\bibinfo {author} {\bibfnamefont {I.~D.}\ \bibnamefont
  {Gialamas}}, \bibinfo {author} {\bibfnamefont {T.}~\bibnamefont
  {Katsoulas}},\ and\ \bibinfo {author} {\bibfnamefont {K.}~\bibnamefont
  {Tamvakis}},\ }\bibfield  {title} {\bibinfo {title} {{Keeping the relation
  between the Starobinsky model and no-scale supergravity ACTive}},\
  }\href@noop {} {\  (\bibinfo {year} {2025}{\natexlab{b}})},\ \Eprint
  {https://arxiv.org/abs/2505.03608} {arXiv:2505.03608 [gr-qc]} \BibitemShut
  {NoStop}%
\bibitem [{\citenamefont {Saini}\ and\ \citenamefont
  {Nautiyal}(2025)}]{Saini:2025jlc}%
  \BibitemOpen
  \bibfield  {author} {\bibinfo {author} {\bibfnamefont {S.}~\bibnamefont
  {Saini}}\ and\ \bibinfo {author} {\bibfnamefont {A.}~\bibnamefont
  {Nautiyal}},\ }\bibfield  {title} {\bibinfo {title} {{Power law
  $\alpha$-Starobinsky inflation}},\ }\href@noop {} {\  (\bibinfo {year}
  {2025})},\ \Eprint {https://arxiv.org/abs/2505.16853} {arXiv:2505.16853
  [astro-ph.CO]} \BibitemShut {NoStop}%
\bibitem [{\citenamefont {Wolf}(2025)}]{Wolf:2025ecy}%
  \BibitemOpen
  \bibfield  {author} {\bibinfo {author} {\bibfnamefont {W.~J.}\ \bibnamefont
  {Wolf}},\ }\bibfield  {title} {\bibinfo {title} {{Inflationary attractors and
  radiative corrections in light of ACT data}},\ }\href@noop {} {\  (\bibinfo
  {year} {2025})},\ \Eprint {https://arxiv.org/abs/2506.12436}
  {arXiv:2506.12436 [astro-ph.CO]} \BibitemShut {NoStop}%
\bibitem [{\citenamefont {Wang}\ \emph {et~al.}(2025)\citenamefont {Wang},
  \citenamefont {Kohri},\ and\ \citenamefont {Yanagida}}]{Wang:2025dbj}%
  \BibitemOpen
  \bibfield  {author} {\bibinfo {author} {\bibfnamefont {X.}~\bibnamefont
  {Wang}}, \bibinfo {author} {\bibfnamefont {K.}~\bibnamefont {Kohri}},\ and\
  \bibinfo {author} {\bibfnamefont {T.~T.}\ \bibnamefont {Yanagida}},\
  }\bibfield  {title} {\bibinfo {title} {{Primordial Black Holes Save $R^2$
  Inflation}},\ }\href@noop {} {\  (\bibinfo {year} {2025})},\ \Eprint
  {https://arxiv.org/abs/2506.06797} {arXiv:2506.06797 [astro-ph.CO]}
  \BibitemShut {NoStop}%
\bibitem [{\citenamefont {Piva}(2025)}]{Piva:2025cqi}%
  \BibitemOpen
  \bibfield  {author} {\bibinfo {author} {\bibfnamefont {M.}~\bibnamefont
  {Piva}},\ }\bibfield  {title} {\bibinfo {title} {{Classification of $f(R)$
  Theories Of Inflation And The Uniqueness of Starobinsky Model}},\ }\href@noop
  {} {\  (\bibinfo {year} {2025})},\ \Eprint {https://arxiv.org/abs/2507.02637}
  {arXiv:2507.02637 [hep-th]} \BibitemShut {NoStop}%
\bibitem [{\citenamefont {Sidik~Risdianto}\ \emph {et~al.}(2025)\citenamefont
  {Sidik~Risdianto}, \citenamefont {Budhi}, \citenamefont {Shobcha},\ and\
  \citenamefont {Salim~Adam}}]{SidikRisdianto:2025qvk}%
  \BibitemOpen
  \bibfield  {author} {\bibinfo {author} {\bibfnamefont {N.}~\bibnamefont
  {Sidik~Risdianto}}, \bibinfo {author} {\bibfnamefont {R.~H.~S.}\ \bibnamefont
  {Budhi}}, \bibinfo {author} {\bibfnamefont {N.}~\bibnamefont {Shobcha}},\
  and\ \bibinfo {author} {\bibfnamefont {A.}~\bibnamefont {Salim~Adam}},\
  }\bibfield  {title} {\bibinfo {title} {{The Preheating Stage on The
  Starobinsky Inflation after ACT}},\ }\href@noop {} {\  (\bibinfo {year}
  {2025})},\ \Eprint {https://arxiv.org/abs/2507.12868} {arXiv:2507.12868
  [gr-qc]} \BibitemShut {NoStop}%
\bibitem [{\citenamefont {Zharov}\ \emph
  {et~al.}(2025{\natexlab{b}})\citenamefont {Zharov}, \citenamefont {Sobol},\
  and\ \citenamefont {Vilchinskii}}]{Zharov:2025zjg}%
  \BibitemOpen
  \bibfield  {author} {\bibinfo {author} {\bibfnamefont {D.~S.}\ \bibnamefont
  {Zharov}}, \bibinfo {author} {\bibfnamefont {O.~O.}\ \bibnamefont {Sobol}},\
  and\ \bibinfo {author} {\bibfnamefont {S.~I.}\ \bibnamefont {Vilchinskii}},\
  }\bibfield  {title} {\bibinfo {title} {{ACT observations, reheating, and
  Starobinsky and Higgs inflation}},\ }\href
  {https://doi.org/10.1103/km3q-rm34} {\bibfield  {journal} {\bibinfo
  {journal} {Phys. Rev. D}\ }\textbf {\bibinfo {volume} {112}},\ \bibinfo
  {pages} {023544} (\bibinfo {year} {2025}{\natexlab{b}})}\BibitemShut
  {NoStop}%
\bibitem [{\citenamefont {Ferreira}\ \emph {et~al.}(2025)\citenamefont
  {Ferreira}, \citenamefont {McDonough}, \citenamefont {Balkenhol},
  \citenamefont {Kallosh}, \citenamefont {Knox},\ and\ \citenamefont
  {Linde}}]{Ferreira:2025lrd}%
  \BibitemOpen
  \bibfield  {author} {\bibinfo {author} {\bibfnamefont {E.~G.~M.}\
  \bibnamefont {Ferreira}}, \bibinfo {author} {\bibfnamefont {E.}~\bibnamefont
  {McDonough}}, \bibinfo {author} {\bibfnamefont {L.}~\bibnamefont
  {Balkenhol}}, \bibinfo {author} {\bibfnamefont {R.}~\bibnamefont {Kallosh}},
  \bibinfo {author} {\bibfnamefont {L.}~\bibnamefont {Knox}},\ and\ \bibinfo
  {author} {\bibfnamefont {A.}~\bibnamefont {Linde}},\ }\bibfield  {title}
  {\bibinfo {title} {{The BAO-CMB Tension and Implications for Inflation}},\
  }\href@noop {} {\  (\bibinfo {year} {2025})},\ \Eprint
  {https://arxiv.org/abs/2507.12459} {arXiv:2507.12459 [astro-ph.CO]}
  \BibitemShut {NoStop}%
\bibitem [{\citenamefont {Ellis}\ \emph {et~al.}(2025)\citenamefont {Ellis},
  \citenamefont {Garcia}, \citenamefont {Nagata}, \citenamefont {Nanopoulos},\
  and\ \citenamefont {Olive}}]{Ellis:2025ieh}%
  \BibitemOpen
  \bibfield  {author} {\bibinfo {author} {\bibfnamefont {J.}~\bibnamefont
  {Ellis}}, \bibinfo {author} {\bibfnamefont {M.~A.~G.}\ \bibnamefont
  {Garcia}}, \bibinfo {author} {\bibfnamefont {N.}~\bibnamefont {Nagata}},
  \bibinfo {author} {\bibfnamefont {D.~V.}\ \bibnamefont {Nanopoulos}},\ and\
  \bibinfo {author} {\bibfnamefont {K.~A.}\ \bibnamefont {Olive}},\ }\bibfield
  {title} {\bibinfo {title} {{Deformations of Starobinsky Inflation in No-Scale
  SU(5) and SO(10) GUTs}},\ }\href@noop {} {\  (\bibinfo {year} {2025})},\
  \Eprint {https://arxiv.org/abs/2508.13279} {arXiv:2508.13279 [hep-ph]}
  \BibitemShut {NoStop}%
\bibitem [{\citenamefont {Oikonomou}(2025)}]{Oikonomou:2025htz}%
  \BibitemOpen
  \bibfield  {author} {\bibinfo {author} {\bibfnamefont {V.~K.}\ \bibnamefont
  {Oikonomou}},\ }\bibfield  {title} {\bibinfo {title} {{Strong Gravity Effects
  on $\mathcal{R}^2$-corrected Single Scalar Field Inflation and Compatibility
  with the ACT Data}},\ }\href@noop {} {\  (\bibinfo {year} {2025})},\ \Eprint
  {https://arxiv.org/abs/2508.17363} {arXiv:2508.17363 [gr-qc]} \BibitemShut
  {NoStop}%
\bibitem [{\citenamefont {Modak}\ \emph {et~al.}(2023)\citenamefont {Modak},
  \citenamefont {R{\"o}ver}, \citenamefont {Sch{\"a}fer}, \citenamefont
  {Schosser},\ and\ \citenamefont {Plehn}}]{Modak:2022gol}%
  \BibitemOpen
  \bibfield  {author} {\bibinfo {author} {\bibfnamefont {T.}~\bibnamefont
  {Modak}}, \bibinfo {author} {\bibfnamefont {L.}~\bibnamefont {R{\"o}ver}},
  \bibinfo {author} {\bibfnamefont {B.~M.}\ \bibnamefont {Sch{\"a}fer}},
  \bibinfo {author} {\bibfnamefont {B.}~\bibnamefont {Schosser}},\ and\
  \bibinfo {author} {\bibfnamefont {T.}~\bibnamefont {Plehn}},\ }\bibfield
  {title} {\bibinfo {title} {{Cornering extended Starobinsky inflation with CMB
  and SKA}},\ }\href {https://doi.org/10.21468/SciPostPhys.15.2.047} {\bibfield
   {journal} {\bibinfo  {journal} {SciPost Phys.}\ }\textbf {\bibinfo {volume}
  {15}},\ \bibinfo {pages} {047} (\bibinfo {year} {2023})},\ \Eprint
  {https://arxiv.org/abs/2210.05698} {arXiv:2210.05698 [astro-ph.CO]}
  \BibitemShut {NoStop}%
\bibitem [{\citenamefont {Lee}\ \emph {et~al.}(2023)\citenamefont {Lee},
  \citenamefont {Modak}, \citenamefont {Oda},\ and\ \citenamefont
  {Takahashi}}]{Lee:2023wdm}%
  \BibitemOpen
  \bibfield  {author} {\bibinfo {author} {\bibfnamefont {S.~M.}\ \bibnamefont
  {Lee}}, \bibinfo {author} {\bibfnamefont {T.}~\bibnamefont {Modak}}, \bibinfo
  {author} {\bibfnamefont {K.-y.}\ \bibnamefont {Oda}},\ and\ \bibinfo {author}
  {\bibfnamefont {T.}~\bibnamefont {Takahashi}},\ }\bibfield  {title} {\bibinfo
  {title} {{Ultraviolet sensitivity in Higgs-Starobinsky inflation}},\ }\href
  {https://doi.org/10.1088/1475-7516/2023/08/045} {\bibfield  {journal}
  {\bibinfo  {journal} {JCAP}\ }\textbf {\bibinfo {volume} {08}},\ \bibinfo
  {pages} {045}},\ \Eprint {https://arxiv.org/abs/2303.09866} {arXiv:2303.09866
  [hep-ph]} \BibitemShut {NoStop}%
\bibitem [{\citenamefont {Saidov}\ and\ \citenamefont
  {Zhuk}(2010)}]{Saidov:2010wx}%
  \BibitemOpen
  \bibfield  {author} {\bibinfo {author} {\bibfnamefont {T.}~\bibnamefont
  {Saidov}}\ and\ \bibinfo {author} {\bibfnamefont {A.}~\bibnamefont {Zhuk}},\
  }\bibfield  {title} {\bibinfo {title} {{Bouncing inflation in nonlinear
  $R^2+R^4$ gravitational model}},\ }\href
  {https://doi.org/10.1103/PhysRevD.81.124002} {\bibfield  {journal} {\bibinfo
  {journal} {Phys. Rev. D}\ }\textbf {\bibinfo {volume} {81}},\ \bibinfo
  {pages} {124002} (\bibinfo {year} {2010})},\ \Eprint
  {https://arxiv.org/abs/1002.4138} {arXiv:1002.4138 [hep-th]} \BibitemShut
  {NoStop}%
\bibitem [{\citenamefont {Huang}(2014)}]{Huang:2013hsb}%
  \BibitemOpen
  \bibfield  {author} {\bibinfo {author} {\bibfnamefont {Q.-G.}\ \bibnamefont
  {Huang}},\ }\bibfield  {title} {\bibinfo {title} {{A polynomial f(R)
  inflation model}},\ }\href {https://doi.org/10.1088/1475-7516/2014/02/035}
  {\bibfield  {journal} {\bibinfo  {journal} {JCAP}\ }\textbf {\bibinfo
  {volume} {02}},\ \bibinfo {pages} {035}},\ \Eprint
  {https://arxiv.org/abs/1309.3514} {arXiv:1309.3514 [hep-th]} \BibitemShut
  {NoStop}%
\bibitem [{\citenamefont {Motohashi}(2015)}]{Motohashi:2014tra}%
  \BibitemOpen
  \bibfield  {author} {\bibinfo {author} {\bibfnamefont {H.}~\bibnamefont
  {Motohashi}},\ }\bibfield  {title} {\bibinfo {title} {{Consistency relation
  for $R^p$ inflation}},\ }\href {https://doi.org/10.1103/PhysRevD.91.064016}
  {\bibfield  {journal} {\bibinfo  {journal} {Phys. Rev. D}\ }\textbf {\bibinfo
  {volume} {91}},\ \bibinfo {pages} {064016} (\bibinfo {year} {2015})},\
  \Eprint {https://arxiv.org/abs/1411.2972} {arXiv:1411.2972 [astro-ph.CO]}
  \BibitemShut {NoStop}%
\bibitem [{\citenamefont {Asaka}\ \emph {et~al.}(2016)\citenamefont {Asaka},
  \citenamefont {Iso}, \citenamefont {Kawai}, \citenamefont {Kohri},
  \citenamefont {Noumi},\ and\ \citenamefont {Terada}}]{Asaka:2015vza}%
  \BibitemOpen
  \bibfield  {author} {\bibinfo {author} {\bibfnamefont {T.}~\bibnamefont
  {Asaka}}, \bibinfo {author} {\bibfnamefont {S.}~\bibnamefont {Iso}}, \bibinfo
  {author} {\bibfnamefont {H.}~\bibnamefont {Kawai}}, \bibinfo {author}
  {\bibfnamefont {K.}~\bibnamefont {Kohri}}, \bibinfo {author} {\bibfnamefont
  {T.}~\bibnamefont {Noumi}},\ and\ \bibinfo {author} {\bibfnamefont
  {T.}~\bibnamefont {Terada}},\ }\bibfield  {title} {\bibinfo {title}
  {{Reinterpretation of the Starobinsky model}},\ }\href
  {https://doi.org/10.1093/ptep/ptw161} {\bibfield  {journal} {\bibinfo
  {journal} {PTEP}\ }\textbf {\bibinfo {volume} {2016}},\ \bibinfo {pages}
  {123E01} (\bibinfo {year} {2016})},\ \Eprint
  {https://arxiv.org/abs/1507.04344} {arXiv:1507.04344 [hep-th]} \BibitemShut
  {NoStop}%
\bibitem [{\citenamefont {Bamba}\ and\ \citenamefont
  {Odintsov}(2015)}]{Bamba:2015uma}%
  \BibitemOpen
  \bibfield  {author} {\bibinfo {author} {\bibfnamefont {K.}~\bibnamefont
  {Bamba}}\ and\ \bibinfo {author} {\bibfnamefont {S.~D.}\ \bibnamefont
  {Odintsov}},\ }\bibfield  {title} {\bibinfo {title} {{Inflationary cosmology
  in modified gravity theories}},\ }\href {https://doi.org/10.3390/sym7010220}
  {\bibfield  {journal} {\bibinfo  {journal} {Symmetry}\ }\textbf {\bibinfo
  {volume} {7}},\ \bibinfo {pages} {220} (\bibinfo {year} {2015})},\ \Eprint
  {https://arxiv.org/abs/1503.00442} {arXiv:1503.00442 [hep-th]} \BibitemShut
  {NoStop}%
\bibitem [{\citenamefont {Miranda}\ \emph {et~al.}(2017)\citenamefont
  {Miranda}, \citenamefont {Fabris},\ and\ \citenamefont
  {Piattella}}]{Miranda:2017juz}%
  \BibitemOpen
  \bibfield  {author} {\bibinfo {author} {\bibfnamefont {T.}~\bibnamefont
  {Miranda}}, \bibinfo {author} {\bibfnamefont {J.~C.}\ \bibnamefont
  {Fabris}},\ and\ \bibinfo {author} {\bibfnamefont {O.~F.}\ \bibnamefont
  {Piattella}},\ }\bibfield  {title} {\bibinfo {title} {{Reconstructing a
  $f(R)$ theory from the $\alpha$-Attractors}},\ }\href
  {https://doi.org/10.1088/1475-7516/2017/09/041} {\bibfield  {journal}
  {\bibinfo  {journal} {JCAP}\ }\textbf {\bibinfo {volume} {09}},\ \bibinfo
  {pages} {041}},\ \Eprint {https://arxiv.org/abs/1707.06457} {arXiv:1707.06457
  [gr-qc]} \BibitemShut {NoStop}%
\bibitem [{\citenamefont {Cheong}\ \emph {et~al.}(2020)\citenamefont {Cheong},
  \citenamefont {Lee},\ and\ \citenamefont {Park}}]{Cheong:2020rao}%
  \BibitemOpen
  \bibfield  {author} {\bibinfo {author} {\bibfnamefont {D.~Y.}\ \bibnamefont
  {Cheong}}, \bibinfo {author} {\bibfnamefont {H.~M.}\ \bibnamefont {Lee}},\
  and\ \bibinfo {author} {\bibfnamefont {S.~C.}\ \bibnamefont {Park}},\
  }\bibfield  {title} {\bibinfo {title} {{Beyond the Starobinsky model for
  inflation}},\ }\href {https://doi.org/10.1016/j.physletb.2020.135453}
  {\bibfield  {journal} {\bibinfo  {journal} {Phys. Lett. B}\ }\textbf
  {\bibinfo {volume} {805}},\ \bibinfo {pages} {135453} (\bibinfo {year}
  {2020})},\ \Eprint {https://arxiv.org/abs/2002.07981} {arXiv:2002.07981
  [hep-ph]} \BibitemShut {NoStop}%
\bibitem [{\citenamefont {Rodrigues-da Silva}\ \emph
  {et~al.}(2022)\citenamefont {Rodrigues-da Silva}, \citenamefont
  {Bezerra-Sobrinho},\ and\ \citenamefont
  {Medeiros}}]{Rodrigues-da-Silva:2021jab}%
  \BibitemOpen
  \bibfield  {author} {\bibinfo {author} {\bibfnamefont {G.}~\bibnamefont
  {Rodrigues-da Silva}}, \bibinfo {author} {\bibfnamefont {J.}~\bibnamefont
  {Bezerra-Sobrinho}},\ and\ \bibinfo {author} {\bibfnamefont {L.~G.}\
  \bibnamefont {Medeiros}},\ }\bibfield  {title} {\bibinfo {title}
  {{Higher-order extension of Starobinsky inflation: Initial conditions,
  slow-roll regime, and reheating phase}},\ }\href
  {https://doi.org/10.1103/PhysRevD.105.063504} {\bibfield  {journal} {\bibinfo
   {journal} {Phys. Rev. D}\ }\textbf {\bibinfo {volume} {105}},\ \bibinfo
  {pages} {063504} (\bibinfo {year} {2022})},\ \Eprint
  {https://arxiv.org/abs/2110.15502} {arXiv:2110.15502 [astro-ph.CO]}
  \BibitemShut {NoStop}%
\bibitem [{\citenamefont {Ivanov}\ \emph {et~al.}(2022)\citenamefont {Ivanov},
  \citenamefont {Ketov}, \citenamefont {Pozdeeva},\ and\ \citenamefont
  {Vernov}}]{Ivanov:2021chn}%
  \BibitemOpen
  \bibfield  {author} {\bibinfo {author} {\bibfnamefont {V.~R.}\ \bibnamefont
  {Ivanov}}, \bibinfo {author} {\bibfnamefont {S.~V.}\ \bibnamefont {Ketov}},
  \bibinfo {author} {\bibfnamefont {E.~O.}\ \bibnamefont {Pozdeeva}},\ and\
  \bibinfo {author} {\bibfnamefont {S.~Y.}\ \bibnamefont {Vernov}},\ }\bibfield
   {title} {\bibinfo {title} {{Analytic extensions of Starobinsky model of
  inflation}},\ }\href {https://doi.org/10.1088/1475-7516/2022/03/058}
  {\bibfield  {journal} {\bibinfo  {journal} {JCAP}\ }\textbf {\bibinfo
  {volume} {03}}\bibfield  {number} {\bibinfo  {number} { (03)},\ \bibinfo
  {pages} {058}},\ }\Eprint {https://arxiv.org/abs/2111.09058}
  {arXiv:2111.09058 [gr-qc]} \BibitemShut {NoStop}%
\bibitem [{\citenamefont {Koshelev}\ \emph {et~al.}(2023)\citenamefont
  {Koshelev}, \citenamefont {Kumar},\ and\ \citenamefont
  {Starobinsky}}]{Koshelev:2022olc}%
  \BibitemOpen
  \bibfield  {author} {\bibinfo {author} {\bibfnamefont {A.~S.}\ \bibnamefont
  {Koshelev}}, \bibinfo {author} {\bibfnamefont {K.~S.}\ \bibnamefont
  {Kumar}},\ and\ \bibinfo {author} {\bibfnamefont {A.~A.}\ \bibnamefont
  {Starobinsky}},\ }\bibfield  {title} {\bibinfo {title} {{Generalized
  non-local R$^{2}$-like inflation}},\ }\href
  {https://doi.org/10.1007/JHEP07(2023)146} {\bibfield  {journal} {\bibinfo
  {journal} {JHEP}\ }\textbf {\bibinfo {volume} {07}},\ \bibinfo {pages}
  {146}},\ \Eprint {https://arxiv.org/abs/2209.02515} {arXiv:2209.02515
  [hep-th]} \BibitemShut {NoStop}%
\bibitem [{\citenamefont {Shtanov}\ \emph {et~al.}(2023)\citenamefont
  {Shtanov}, \citenamefont {Sahni},\ and\ \citenamefont
  {Mishra}}]{Shtanov:2022pdx}%
  \BibitemOpen
  \bibfield  {author} {\bibinfo {author} {\bibfnamefont {Y.}~\bibnamefont
  {Shtanov}}, \bibinfo {author} {\bibfnamefont {V.}~\bibnamefont {Sahni}},\
  and\ \bibinfo {author} {\bibfnamefont {S.~S.}\ \bibnamefont {Mishra}},\
  }\bibfield  {title} {\bibinfo {title} {{Tabletop potentials for inflation
  from f(R) gravity}},\ }\href {https://doi.org/10.1088/1475-7516/2023/03/023}
  {\bibfield  {journal} {\bibinfo  {journal} {JCAP}\ }\textbf {\bibinfo
  {volume} {03}},\ \bibinfo {pages} {023}},\ \Eprint
  {https://arxiv.org/abs/2210.01828} {arXiv:2210.01828 [gr-qc]} \BibitemShut
  {NoStop}%
\bibitem [{\citenamefont {Wang}\ \emph {et~al.}(2023)\citenamefont {Wang},
  \citenamefont {Tang},\ and\ \citenamefont {Wu}}]{Wang:2023hsb}%
  \BibitemOpen
  \bibfield  {author} {\bibinfo {author} {\bibfnamefont {Q.-Y.}\ \bibnamefont
  {Wang}}, \bibinfo {author} {\bibfnamefont {Y.}~\bibnamefont {Tang}},\ and\
  \bibinfo {author} {\bibfnamefont {Y.-L.}\ \bibnamefont {Wu}},\ }\bibfield
  {title} {\bibinfo {title} {{Inflation in Weyl scaling invariant gravity with
  $R^3$ extensions}},\ }\href {https://doi.org/10.1103/PhysRevD.107.083511}
  {\bibfield  {journal} {\bibinfo  {journal} {Phys. Rev. D}\ }\textbf {\bibinfo
  {volume} {107}},\ \bibinfo {pages} {083511} (\bibinfo {year} {2023})},\
  \Eprint {https://arxiv.org/abs/2301.03744} {arXiv:2301.03744 [astro-ph.CO]}
  \BibitemShut {NoStop}%
\bibitem [{\citenamefont {Kim}\ \emph {et~al.}(2025)\citenamefont {Kim},
  \citenamefont {Wang}, \citenamefont {Zhang},\ and\ \citenamefont
  {Ren}}]{Kim:2025dyi}%
  \BibitemOpen
  \bibfield  {author} {\bibinfo {author} {\bibfnamefont {J.}~\bibnamefont
  {Kim}}, \bibinfo {author} {\bibfnamefont {X.}~\bibnamefont {Wang}}, \bibinfo
  {author} {\bibfnamefont {Y.-l.}\ \bibnamefont {Zhang}},\ and\ \bibinfo
  {author} {\bibfnamefont {Z.}~\bibnamefont {Ren}},\ }\bibfield  {title}
  {\bibinfo {title} {{Enhancement of primordial curvature perturbations in R
  $^{3}$-corrected Starobinsky-Higgs inflation}},\ }\href
  {https://doi.org/10.1088/1475-7516/2025/09/011} {\bibfield  {journal}
  {\bibinfo  {journal} {JCAP}\ }\textbf {\bibinfo {volume} {09}},\ \bibinfo
  {pages} {011}},\ \Eprint {https://arxiv.org/abs/2504.12035} {arXiv:2504.12035
  [astro-ph.CO]} \BibitemShut {NoStop}%
\bibitem [{\citenamefont {Sebastiani}\ \emph {et~al.}(2014)\citenamefont
  {Sebastiani}, \citenamefont {Cognola}, \citenamefont {Myrzakulov},
  \citenamefont {Odintsov},\ and\ \citenamefont
  {Zerbini}}]{Sebastiani:2013eqa}%
  \BibitemOpen
  \bibfield  {author} {\bibinfo {author} {\bibfnamefont {L.}~\bibnamefont
  {Sebastiani}}, \bibinfo {author} {\bibfnamefont {G.}~\bibnamefont {Cognola}},
  \bibinfo {author} {\bibfnamefont {R.}~\bibnamefont {Myrzakulov}}, \bibinfo
  {author} {\bibfnamefont {S.~D.}\ \bibnamefont {Odintsov}},\ and\ \bibinfo
  {author} {\bibfnamefont {S.}~\bibnamefont {Zerbini}},\ }\bibfield  {title}
  {\bibinfo {title} {{Nearly Starobinsky inflation from modified gravity}},\
  }\href {https://doi.org/10.1103/PhysRevD.89.023518} {\bibfield  {journal}
  {\bibinfo  {journal} {Phys. Rev. D}\ }\textbf {\bibinfo {volume} {89}},\
  \bibinfo {pages} {023518} (\bibinfo {year} {2014})},\ \Eprint
  {https://arxiv.org/abs/1311.0744} {arXiv:1311.0744 [gr-qc]} \BibitemShut
  {NoStop}%
\bibitem [{\citenamefont {Odintsov}\ \emph {et~al.}(2017)\citenamefont
  {Odintsov}, \citenamefont {Oikonomou},\ and\ \citenamefont
  {Sebastiani}}]{Odintsov:2017hbk}%
  \BibitemOpen
  \bibfield  {author} {\bibinfo {author} {\bibfnamefont {S.~D.}\ \bibnamefont
  {Odintsov}}, \bibinfo {author} {\bibfnamefont {V.~K.}\ \bibnamefont
  {Oikonomou}},\ and\ \bibinfo {author} {\bibfnamefont {L.}~\bibnamefont
  {Sebastiani}},\ }\bibfield  {title} {\bibinfo {title} {{Unification of
  Constant-roll Inflation and Dark Energy with Logarithmic $R^2$-corrected and
  Exponential $F(R)$ Gravity}},\ }\href
  {https://doi.org/10.1016/j.nuclphysb.2017.08.018} {\bibfield  {journal}
  {\bibinfo  {journal} {Nucl. Phys. B}\ }\textbf {\bibinfo {volume} {923}},\
  \bibinfo {pages} {608} (\bibinfo {year} {2017})},\ \Eprint
  {https://arxiv.org/abs/1708.08346} {arXiv:1708.08346 [gr-qc]} \BibitemShut
  {NoStop}%
\bibitem [{\citenamefont {Nojiri}\ \emph {et~al.}(2017)\citenamefont {Nojiri},
  \citenamefont {Odintsov},\ and\ \citenamefont {Oikonomou}}]{Nojiri:2017ncd}%
  \BibitemOpen
  \bibfield  {author} {\bibinfo {author} {\bibfnamefont {S.}~\bibnamefont
  {Nojiri}}, \bibinfo {author} {\bibfnamefont {S.~D.}\ \bibnamefont
  {Odintsov}},\ and\ \bibinfo {author} {\bibfnamefont {V.~K.}\ \bibnamefont
  {Oikonomou}},\ }\bibfield  {title} {\bibinfo {title} {{Modified Gravity
  Theories on a Nutshell: Inflation, Bounce and Late-time Evolution}},\ }\href
  {https://doi.org/10.1016/j.physrep.2017.06.001} {\bibfield  {journal}
  {\bibinfo  {journal} {Phys. Rept.}\ }\textbf {\bibinfo {volume} {692}},\
  \bibinfo {pages} {1} (\bibinfo {year} {2017})},\ \Eprint
  {https://arxiv.org/abs/1705.11098} {arXiv:1705.11098 [gr-qc]} \BibitemShut
  {NoStop}%
\bibitem [{\citenamefont {Cado}\ \emph {et~al.}(2025)\citenamefont {Cado},
  \citenamefont {Englert}, \citenamefont {Modak},\ and\ \citenamefont
  {Quir{\'o}s}}]{Cado:2024von}%
  \BibitemOpen
  \bibfield  {author} {\bibinfo {author} {\bibfnamefont {Y.}~\bibnamefont
  {Cado}}, \bibinfo {author} {\bibfnamefont {C.}~\bibnamefont {Englert}},
  \bibinfo {author} {\bibfnamefont {T.}~\bibnamefont {Modak}},\ and\ \bibinfo
  {author} {\bibfnamefont {M.}~\bibnamefont {Quir{\'o}s}},\ }\bibfield  {title}
  {\bibinfo {title} {{Implication of preheating on gravity assisted
  baryogenesis in R2-Higgs inflation}},\ }\href
  {https://doi.org/10.1103/PhysRevD.111.023042} {\bibfield  {journal} {\bibinfo
   {journal} {Phys. Rev. D}\ }\textbf {\bibinfo {volume} {111}},\ \bibinfo
  {pages} {023042} (\bibinfo {year} {2025})},\ \Eprint
  {https://arxiv.org/abs/2411.11128} {arXiv:2411.11128 [hep-ph]} \BibitemShut
  {NoStop}%
\bibitem [{\citenamefont {Gong}\ and\ \citenamefont
  {Tanaka}(2011)}]{Gong:2011uw}%
  \BibitemOpen
  \bibfield  {author} {\bibinfo {author} {\bibfnamefont {J.-O.}\ \bibnamefont
  {Gong}}\ and\ \bibinfo {author} {\bibfnamefont {T.}~\bibnamefont {Tanaka}},\
  }\bibfield  {title} {\bibinfo {title} {{A covariant approach to general field
  space metric in multi-field inflation}},\ }\href
  {https://doi.org/10.1088/1475-7516/2012/02/E01} {\bibfield  {journal}
  {\bibinfo  {journal} {JCAP}\ }\textbf {\bibinfo {volume} {03}},\ \bibinfo
  {pages} {015}},\ \bibinfo {note} {[Erratum: JCAP 02, E01 (2012)]},\ \Eprint
  {https://arxiv.org/abs/1101.4809} {arXiv:1101.4809 [astro-ph.CO]}
  \BibitemShut {NoStop}%
\bibitem [{\citenamefont {Sfakianakis}\ and\ \citenamefont {van~de
  Vis}(2019)}]{Sfakianakis:2018lzf}%
  \BibitemOpen
  \bibfield  {author} {\bibinfo {author} {\bibfnamefont {E.~I.}\ \bibnamefont
  {Sfakianakis}}\ and\ \bibinfo {author} {\bibfnamefont {J.}~\bibnamefont
  {van~de Vis}},\ }\bibfield  {title} {\bibinfo {title} {{Preheating after
  Higgs Inflation: Self-Resonance and Gauge boson production}},\ }\href
  {https://doi.org/10.1103/PhysRevD.99.083519} {\bibfield  {journal} {\bibinfo
  {journal} {Phys. Rev. D}\ }\textbf {\bibinfo {volume} {99}},\ \bibinfo
  {pages} {083519} (\bibinfo {year} {2019})},\ \Eprint
  {https://arxiv.org/abs/1810.01304} {arXiv:1810.01304 [hep-ph]} \BibitemShut
  {NoStop}%
\bibitem [{\citenamefont {Kaiser}\ \emph {et~al.}(2013)\citenamefont {Kaiser},
  \citenamefont {Mazenc},\ and\ \citenamefont {Sfakianakis}}]{Kaiser:2012ak}%
  \BibitemOpen
  \bibfield  {author} {\bibinfo {author} {\bibfnamefont {D.~I.}\ \bibnamefont
  {Kaiser}}, \bibinfo {author} {\bibfnamefont {E.~A.}\ \bibnamefont {Mazenc}},\
  and\ \bibinfo {author} {\bibfnamefont {E.~I.}\ \bibnamefont {Sfakianakis}},\
  }\bibfield  {title} {\bibinfo {title} {{Primordial Bispectrum from Multifield
  Inflation with Nonminimal Couplings}},\ }\href
  {https://doi.org/10.1103/PhysRevD.87.064004} {\bibfield  {journal} {\bibinfo
  {journal} {Phys. Rev. D}\ }\textbf {\bibinfo {volume} {87}},\ \bibinfo
  {pages} {064004} (\bibinfo {year} {2013})},\ \Eprint
  {https://arxiv.org/abs/1210.7487} {arXiv:1210.7487 [astro-ph.CO]}
  \BibitemShut {NoStop}%
\bibitem [{\citenamefont {Kodama}\ and\ \citenamefont
  {Sasaki}(1984)}]{Kodama:1984ziu}%
  \BibitemOpen
  \bibfield  {author} {\bibinfo {author} {\bibfnamefont {H.}~\bibnamefont
  {Kodama}}\ and\ \bibinfo {author} {\bibfnamefont {M.}~\bibnamefont
  {Sasaki}},\ }\bibfield  {title} {\bibinfo {title} {{Cosmological Perturbation
  Theory}},\ }\href {https://doi.org/10.1143/PTPS.78.1} {\bibfield  {journal}
  {\bibinfo  {journal} {Prog. Theor. Phys. Suppl.}\ }\textbf {\bibinfo {volume}
  {78}},\ \bibinfo {pages} {1} (\bibinfo {year} {1984})}\BibitemShut {NoStop}%
\bibitem [{\citenamefont {Mukhanov}\ \emph {et~al.}(1992)\citenamefont
  {Mukhanov}, \citenamefont {Feldman},\ and\ \citenamefont
  {Brandenberger}}]{Mukhanov:1990me}%
  \BibitemOpen
  \bibfield  {author} {\bibinfo {author} {\bibfnamefont {V.~F.}\ \bibnamefont
  {Mukhanov}}, \bibinfo {author} {\bibfnamefont {H.~A.}\ \bibnamefont
  {Feldman}},\ and\ \bibinfo {author} {\bibfnamefont {R.~H.}\ \bibnamefont
  {Brandenberger}},\ }\bibfield  {title} {\bibinfo {title} {{Theory of
  cosmological perturbations. Part 1. Classical perturbations. Part 2. Quantum
  theory of perturbations. Part 3. Extensions}},\ }\href
  {https://doi.org/10.1016/0370-1573(92)90044-Z} {\bibfield  {journal}
  {\bibinfo  {journal} {Phys. Rept.}\ }\textbf {\bibinfo {volume} {215}},\
  \bibinfo {pages} {203} (\bibinfo {year} {1992})}\BibitemShut {NoStop}%
\bibitem [{\citenamefont {Malik}\ and\ \citenamefont
  {Wands}(2009)}]{Malik:2008im}%
  \BibitemOpen
  \bibfield  {author} {\bibinfo {author} {\bibfnamefont {K.~A.}\ \bibnamefont
  {Malik}}\ and\ \bibinfo {author} {\bibfnamefont {D.}~\bibnamefont {Wands}},\
  }\bibfield  {title} {\bibinfo {title} {{Cosmological perturbations}},\ }\href
  {https://doi.org/10.1016/j.physrep.2009.03.001} {\bibfield  {journal}
  {\bibinfo  {journal} {Phys. Rept.}\ }\textbf {\bibinfo {volume} {475}},\
  \bibinfo {pages} {1} (\bibinfo {year} {2009})},\ \Eprint
  {https://arxiv.org/abs/0809.4944} {arXiv:0809.4944 [astro-ph]} \BibitemShut
  {NoStop}%
\bibitem [{\citenamefont {Elliston}\ \emph {et~al.}(2012)\citenamefont
  {Elliston}, \citenamefont {Seery},\ and\ \citenamefont
  {Tavakol}}]{Elliston:2012ab}%
  \BibitemOpen
  \bibfield  {author} {\bibinfo {author} {\bibfnamefont {J.}~\bibnamefont
  {Elliston}}, \bibinfo {author} {\bibfnamefont {D.}~\bibnamefont {Seery}},\
  and\ \bibinfo {author} {\bibfnamefont {R.}~\bibnamefont {Tavakol}},\
  }\bibfield  {title} {\bibinfo {title} {{The inflationary bispectrum with
  curved field-space}},\ }\href {https://doi.org/10.1088/1475-7516/2012/11/060}
  {\bibfield  {journal} {\bibinfo  {journal} {JCAP}\ }\textbf {\bibinfo
  {volume} {11}},\ \bibinfo {pages} {060}},\ \Eprint
  {https://arxiv.org/abs/1208.6011} {arXiv:1208.6011 [astro-ph.CO]}
  \BibitemShut {NoStop}%
\bibitem [{\citenamefont {Sasaki}(1986)}]{Sasaki:1986hm}%
  \BibitemOpen
  \bibfield  {author} {\bibinfo {author} {\bibfnamefont {M.}~\bibnamefont
  {Sasaki}},\ }\bibfield  {title} {\bibinfo {title} {{Large Scale Quantum
  Fluctuations in the Inflationary Universe}},\ }\href
  {https://doi.org/10.1143/PTP.76.1036} {\bibfield  {journal} {\bibinfo
  {journal} {Prog. Theor. Phys.}\ }\textbf {\bibinfo {volume} {76}},\ \bibinfo
  {pages} {1036} (\bibinfo {year} {1986})}\BibitemShut {NoStop}%
\bibitem [{\citenamefont {Mukhanov}(1988)}]{Mukhanov:1988jd}%
  \BibitemOpen
  \bibfield  {author} {\bibinfo {author} {\bibfnamefont {V.~F.}\ \bibnamefont
  {Mukhanov}},\ }\bibfield  {title} {\bibinfo {title} {{Quantum Theory of Gauge
  Invariant Cosmological Perturbations}},\ }\href@noop {} {\bibfield  {journal}
  {\bibinfo  {journal} {Sov. Phys. JETP}\ }\textbf {\bibinfo {volume} {67}},\
  \bibinfo {pages} {1297} (\bibinfo {year} {1988})}\BibitemShut {NoStop}%
\bibitem [{\citenamefont {Bassett}\ \emph {et~al.}(2006)\citenamefont
  {Bassett}, \citenamefont {Tsujikawa},\ and\ \citenamefont
  {Wands}}]{Bassett:2005xm}%
  \BibitemOpen
  \bibfield  {author} {\bibinfo {author} {\bibfnamefont {B.~A.}\ \bibnamefont
  {Bassett}}, \bibinfo {author} {\bibfnamefont {S.}~\bibnamefont {Tsujikawa}},\
  and\ \bibinfo {author} {\bibfnamefont {D.}~\bibnamefont {Wands}},\ }\bibfield
   {title} {\bibinfo {title} {{Inflation dynamics and reheating}},\ }\href
  {https://doi.org/10.1103/RevModPhys.78.537} {\bibfield  {journal} {\bibinfo
  {journal} {Rev. Mod. Phys.}\ }\textbf {\bibinfo {volume} {78}},\ \bibinfo
  {pages} {537} (\bibinfo {year} {2006})},\ \Eprint
  {https://arxiv.org/abs/astro-ph/0507632} {arXiv:astro-ph/0507632}
  \BibitemShut {NoStop}%
\bibitem [{\citenamefont {Antusch}\ \emph {et~al.}(2015)\citenamefont
  {Antusch}, \citenamefont {Nolde},\ and\ \citenamefont
  {Orani}}]{Antusch:2015nla}%
  \BibitemOpen
  \bibfield  {author} {\bibinfo {author} {\bibfnamefont {S.}~\bibnamefont
  {Antusch}}, \bibinfo {author} {\bibfnamefont {D.}~\bibnamefont {Nolde}},\
  and\ \bibinfo {author} {\bibfnamefont {S.}~\bibnamefont {Orani}},\ }\bibfield
   {title} {\bibinfo {title} {{Hill crossing during preheating after hilltop
  inflation}},\ }\href {https://doi.org/10.1088/1475-7516/2015/06/009}
  {\bibfield  {journal} {\bibinfo  {journal} {JCAP}\ }\textbf {\bibinfo
  {volume} {06}},\ \bibinfo {pages} {009}},\ \Eprint
  {https://arxiv.org/abs/1503.06075} {arXiv:1503.06075 [hep-ph]} \BibitemShut
  {NoStop}%
\bibitem [{\citenamefont {DeCross}\ \emph
  {et~al.}(2018{\natexlab{a}})\citenamefont {DeCross}, \citenamefont {Kaiser},
  \citenamefont {Prabhu}, \citenamefont {Prescod-Weinstein},\ and\
  \citenamefont {Sfakianakis}}]{DeCross:2015uza}%
  \BibitemOpen
  \bibfield  {author} {\bibinfo {author} {\bibfnamefont {M.~P.}\ \bibnamefont
  {DeCross}}, \bibinfo {author} {\bibfnamefont {D.~I.}\ \bibnamefont {Kaiser}},
  \bibinfo {author} {\bibfnamefont {A.}~\bibnamefont {Prabhu}}, \bibinfo
  {author} {\bibfnamefont {C.}~\bibnamefont {Prescod-Weinstein}},\ and\
  \bibinfo {author} {\bibfnamefont {E.~I.}\ \bibnamefont {Sfakianakis}},\
  }\bibfield  {title} {\bibinfo {title} {{Preheating after Multifield Inflation
  with Nonminimal Couplings, I: Covariant Formalism and Attractor Behavior}},\
  }\href {https://doi.org/10.1103/PhysRevD.97.023526} {\bibfield  {journal}
  {\bibinfo  {journal} {Phys. Rev. D}\ }\textbf {\bibinfo {volume} {97}},\
  \bibinfo {pages} {023526} (\bibinfo {year} {2018}{\natexlab{a}})},\ \Eprint
  {https://arxiv.org/abs/1510.08553} {arXiv:1510.08553 [astro-ph.CO]}
  \BibitemShut {NoStop}%
\bibitem [{\citenamefont {DeCross}\ \emph
  {et~al.}(2018{\natexlab{b}})\citenamefont {DeCross}, \citenamefont {Kaiser},
  \citenamefont {Prabhu}, \citenamefont {Prescod-Weinstein},\ and\
  \citenamefont {Sfakianakis}}]{DeCross:2016cbs}%
  \BibitemOpen
  \bibfield  {author} {\bibinfo {author} {\bibfnamefont {M.~P.}\ \bibnamefont
  {DeCross}}, \bibinfo {author} {\bibfnamefont {D.~I.}\ \bibnamefont {Kaiser}},
  \bibinfo {author} {\bibfnamefont {A.}~\bibnamefont {Prabhu}}, \bibinfo
  {author} {\bibfnamefont {C.}~\bibnamefont {Prescod-Weinstein}},\ and\
  \bibinfo {author} {\bibfnamefont {E.~I.}\ \bibnamefont {Sfakianakis}},\
  }\bibfield  {title} {\bibinfo {title} {{Preheating after multifield inflation
  with nonminimal couplings, III: Dynamical spacetime results}},\ }\href
  {https://doi.org/10.1103/PhysRevD.97.023528} {\bibfield  {journal} {\bibinfo
  {journal} {Phys. Rev. D}\ }\textbf {\bibinfo {volume} {97}},\ \bibinfo
  {pages} {023528} (\bibinfo {year} {2018}{\natexlab{b}})},\ \Eprint
  {https://arxiv.org/abs/1610.08916} {arXiv:1610.08916 [astro-ph.CO]}
  \BibitemShut {NoStop}%
\bibitem [{\citenamefont {Cado}\ \emph {et~al.}(2024)\citenamefont {Cado},
  \citenamefont {Englert}, \citenamefont {Modak},\ and\ \citenamefont
  {Quir{\'o}s}}]{Cado:2023zbm}%
  \BibitemOpen
  \bibfield  {author} {\bibinfo {author} {\bibfnamefont {Y.}~\bibnamefont
  {Cado}}, \bibinfo {author} {\bibfnamefont {C.}~\bibnamefont {Englert}},
  \bibinfo {author} {\bibfnamefont {T.}~\bibnamefont {Modak}},\ and\ \bibinfo
  {author} {\bibfnamefont {M.}~\bibnamefont {Quir{\'o}s}},\ }\bibfield  {title}
  {\bibinfo {title} {{Baryogenesis in $R^2$-Higgs inflation: The gravitational
  connection}},\ }\href {https://doi.org/10.1103/PhysRevD.109.043026}
  {\bibfield  {journal} {\bibinfo  {journal} {Phys. Rev. D}\ }\textbf {\bibinfo
  {volume} {109}},\ \bibinfo {pages} {043026} (\bibinfo {year} {2024})},\
  \Eprint {https://arxiv.org/abs/2312.10414} {arXiv:2312.10414 [astro-ph.CO]}
  \BibitemShut {NoStop}%
\bibitem [{\citenamefont {Abitbol}\ \emph {et~al.}(2025)\citenamefont {Abitbol}
  \emph {et~al.}}]{SimonsObservatory:2025wwn}%
  \BibitemOpen
  \bibfield  {author} {\bibinfo {author} {\bibfnamefont {M.}~\bibnamefont
  {Abitbol}} \emph {et~al.} (\bibinfo {collaboration} {Simons Observatory}),\
  }\bibfield  {title} {\bibinfo {title} {{The Simons Observatory: science goals
  and forecasts for the enhanced Large Aperture Telescope}},\ }\href
  {https://doi.org/10.1088/1475-7516/2025/08/034} {\bibfield  {journal}
  {\bibinfo  {journal} {JCAP}\ }\textbf {\bibinfo {volume} {08}},\ \bibinfo
  {pages} {034}},\ \Eprint {https://arxiv.org/abs/2503.00636} {arXiv:2503.00636
  [astro-ph.IM]} \BibitemShut {NoStop}%
\bibitem [{\citenamefont {Abbott}\ \emph {et~al.}(2024)\citenamefont {Abbott}
  \emph {et~al.}}]{DES:2024pwq}%
  \BibitemOpen
  \bibfield  {author} {\bibinfo {author} {\bibfnamefont {T.~M.~C.}\
  \bibnamefont {Abbott}} \emph {et~al.} (\bibinfo {collaboration} {DES}),\
  }\bibfield  {title} {\bibinfo {title} {{Dark Energy Survey: A 2.1{\%}
  measurement of the angular baryonic acoustic oscillation scale at redshift
  $z_{\rm eff}=0.85$ from the final dataset}},\ }\href
  {https://doi.org/10.1103/PhysRevD.110.063515} {\bibfield  {journal} {\bibinfo
   {journal} {Phys. Rev. D}\ }\textbf {\bibinfo {volume} {110}},\ \bibinfo
  {pages} {063515} (\bibinfo {year} {2024})},\ \Eprint
  {https://arxiv.org/abs/2402.10696} {arXiv:2402.10696 [astro-ph.CO]}
  \BibitemShut {NoStop}%
\bibitem [{\citenamefont {Duret}\ \emph {et~al.}(2025)\citenamefont {Duret}
  \emph {et~al.}}]{Euclid:2025dlg}%
  \BibitemOpen
  \bibfield  {author} {\bibinfo {author} {\bibfnamefont {V.}~\bibnamefont
  {Duret}} \emph {et~al.} (\bibinfo {collaboration} {Euclid}),\ }\bibfield
  {title} {\bibinfo {title} {{Euclid preparation. BAO analysis of photometric
  galaxy clustering in configuration space}},\ }\href@noop {} {\  (\bibinfo
  {year} {2025})},\ \Eprint {https://arxiv.org/abs/2503.11621}
  {arXiv:2503.11621 [astro-ph.CO]} \BibitemShut {NoStop}%
\end{thebibliography}%

\end{document}